\documentclass{article}
\usepackage[nolist]{acronym}
\usepackage{amsmath,bm,amsfonts}
\usepackage{graphicx}
\usepackage{appendix}
\usepackage{tikz}
\usetikzlibrary{shapes.geometric, arrows}
\usepackage[section]{placeins}
\usepackage{enumitem}
\usepackage{hyperref}
\usepackage{capekCommands}
\usepackage{mtbpars}
\usepackage{physics}
\usepackage{authblk}
\usepackage{caption}
\usepackage{subcaption}
\usepackage{float}

\hypersetup{
	colorlinks=true, 
	linkcolor=blue!70!black, 
	citecolor=red!70!black, 
	filecolor=blue!70!black, 
	urlcolor=blue!70!black, 
	}

\floatstyle{plain}
\newfloat{program}{thp}{lop}
\floatname{program}{Program}

\newcommand{\MATLAB}{$\mathrm{MATLAB}^\circledR$ }
\newcommand{\expn}[1]{\mathrm{e}^{#1}}

\DeclareMathOperator*{\argmin}{arg\,min}

\newacro{RWG}{Rao-Wilton-Glisson}
\newacro{GEP}{generalized eigenvalue problem}
\newacro{MoM}{method of moments}
\newacro{QCQP}{quadratic constrained quadratic program}
\newacro{AToM}{Antenna Toolbox for \MATLAB}
\newacro{EFIE}{electric field integral equation}

\begin{document}

%
\title{Fundamental Bounds to Time-Harmonic Quadratic Metrics in Electromagnetism: Overview and Implementation}
\author{Jakub Liska\thanks{\href{mailto:jakub.liska@fel.cvut.cz}{jakub.liska@fel.cvut.cz}}}
\author{Lukas Jelinek\thanks{\href{mailto:lukas.jelinek@fel.cvut.cz}{lukas.jelinek@fel.cvut.cz}}}
\author{Miloslav Capek\thanks{\href{mailto:miloslav.capek@fel.cvut.cz}{miloslav.capek@fel.cvut.cz}}}
\affil{Department of Electromagnetic Field,
Faculty of Electrical Engineering,
Czech Technical University in Prague}
\date{December 2021}

\maketitle


\begin{abstract}
\noindent
Fundamental bounds on quadratic electromagnetic metrics are formulated and solved via convex optimization. Both dual formulation and method-of-moments formulation of the electric field integral equation are used as key ingredients. The trade-off between metrics is formulated as a multi-objective optimization resulting in Pareto-optimal sets. Substructure fundamental bounds are also introduced and formulated as additional affine constraints. The general methodology is demonstrated on a few examples of minimal complexity and all examples are supported with freely available MATLAB codes contained in the developed package on fundamental bounds. 
\end{abstract}

\sloppy 
\section{Introduction}
Considerable computational resources are often spent in the topological optimization of electromagnetic structures~\cite{2018_Molesky_NatP}, such as radiators~\cite{Capeketal_ShapeSynthesisBasedOnTopologySensitivity}, waveguide components~\cite{Garcia_OptimizationOfPlanarDevicesByTheFiniteElementMethod}, or cavities~\cite{Walsh_ResonantCavityDesignUsingTheFiniteElementMethod}. Nevertheless, the resulting designs are rarely compared to their physical bounds, i.e., those bounds which provide the best possible metric realization of a given problem~\cite{GustafssonTayliEhrenborgEtAl_AntennaCurrentOptimizationUsingMatlabAndCVX,GustafssonTayliCismasu_PhysicalBoundsOfAntennasBook,JelinekCapek_OptimalCurrentsOnArbitrarilyShapedSurfaces,Capek_etal_2019_OptimalPlanarElectricDipoleAntennas,Skrivervik_Bosiljevac_Sipus_FundamentalBoundsForImplantedAntennas,2020_Molesky_PRR, 2020_Venkataram_PRL, 2020_Kuang_PRL,2021_Chao_Arxiv} and which could be used to judge the performance of the optimizer.

Fundamental bounds in electromagnetism can be typically formulated as \acfp{QCQP}. This allows us to use the tools of convex optimization~\cite{BoydVandenberghe_ConvexOptimization}, a special class of mathematical optimization problems considered to be one of the most employable theories in computational mathematics~\cite{BoydVandenberghe_ConvexOptimization}. Notable examples in this area of study are fundamental bounds to Q-factor~\cite{CapekGustafssonSchab_MinimizationOfAntennaQualityFactor}, radiation efficiency~\cite{GustafssonCapek_MaximumGainEffAreaAndDirectivity, GustafssonCapekSchab_TradeOffBetweenAntennaEfficiencyAndQfactor,Jelinek+etal2018}, antenna gain~\cite{Harrington_AntennaExcitationForMaximumGain,Gustafsson_OptimalAntennaCurrentsForQsuperdirectivityAndRP,GustafssonCapek_MaximumGainEffAreaAndDirectivity}, thermal radiation~\cite{2020_Venkataram_PRL,2020_Molesky_PRB} scattering cross-sections~\cite{2020_Gustafsson_NJP,2020_Molesky_PRR,2021_Jelinek_OPEX} and their trade-offs~\cite{2020_Schab_TradeoffsInAbsAndScatt}.

Applying \ac{MoM} to the electric field integral equation~\cite{Harrington_FieldComputationByMoM} makes it an indispensable tool for establishing fundamental bounds in electromagnetism. In this form, the fundamental bounds can be based on current density expansion into a set of basis functions~\cite{Gibson_MoMinElectromagnetics,Jin_TheoryAndComputationOfElectromagneticFields, RaoWiltonGlisson_ElectromagneticScatteringBySurfacesOfArbitraryShape, Harrington_FieldComputationByMoM} which allows common electromagnetic functionals to be recast as linear or quadratic functions of expansion coefficients. Formulations of fundamental bounds then become \acp{QCQP}.

The aim of this paper is to show that there are many cases where the evaluation of fundamental bounds becomes a routine task which should always be performed to increase the likelihood of success in topological optimization and final design. This evaluation routine necessarily involves computational codes which are presented throughout this paper, together with the links to their web repository.
\section{Time-Harmonic Metrics in Electromagnetism}
Within the \ac{MoM} paradigm~\cite{Harrington_MatrixMethodsForFieldProblems,Harrington_FieldComputationByMoM}, linear electromagnetic operators~$\mathcal{L} \left( \bm{J}(\bm{r}) \right)$ acting on equivalent current density~$\bm{J}(\bm{r})$ are projected onto a set of basis functions $\left\{ \bm{\psi}_n (\bm{r}) \right\}$ by defining
\begin{equation} \label{eq:currDenExpn}
    \bm{J} (\bm{r}, \omega) \approx \sum \limits_{n=1}^N I_n(\omega) \, \basisFcn_n (\bm{r}),
\end{equation}
where the expansion coefficients~$I_n$ are collected in column matrix $\M{I}$. A notable example of an electromagnetic operator is the electric field operator~\cite{Harrington_FieldComputationByMoM,Harrington_AntennaExcitationForMaximumGain}, which, with the help of Poynting's theorem~\cite{Jackson_ClassicalElectrodynamics, Zangwill_Modern_Electrodynamics}, can be used to define power metrics such as dissipated power, reactive power, and radiated power. The procedure detailed in Appendix~\ref{ap:matricesELMAG} shows that most steady-state electromagnetic metrics used can be represented via a quadratic functional
\begin{equation}\label{eq:guadFcnl}
    f(\Ivec) = \Ivec^\herm \M{A} \Ivec + \RE[\Ivec^\herm \M{a}] + \alpha,
\end{equation}
where $^\herm$ denotes the Hermitian conjugate, $\M{A}$ a square Hermitian matrix, $\M{a}$ a column matrix, and $\alpha$ a real constant.

Fundamental bounds are formed from~\eqref{eq:guadFcnl} using expansion coefficients~$I_n$ as degrees of freedom for the optimization where quadratic forms~\eqref{eq:guadFcnl} act as optimized metrics or constraints. The resulting optimization problem falls into the \ac{QCQP} family which is briefly described in Appendix~\ref{ap:QCQP}.
\section{Examples}\label{sec:examples}
This section provides three examples of how fundamental bounds are evaluated. The first example shows a single-objective optimization applied to the lower bound on the radiation Q-factor. The second example concerns multi-objective optimization and shows a three-dimensional Pareto frontier of radiation Q-factor, dissipation factor, and directivity. The third example considers the case of maximum absorption and shows the treatment of external excitation, substructure bounds and the general treatment of affine constraints (connected, in this case, to far-field constraints and partial control). The examples follow the steps in Appendix~\ref{ap:flowChart} which explains the entire procedure of determining the fundamental bounds by the prepared algorithms. The examples are supplemented with several other appendices containing mathematical and implementation details. 

\subsection{Lower Bound on Q-Factor}\label{sec:minQ}
As an example of a single-objective bound, assume that a lower bound on the radiation Q-factor \cite{GustafssonTayliEhrenborgEtAl_AntennaCurrentOptimizationUsingMatlabAndCVX,JelinekCapek_OptimalCurrentsOnArbitrarilyShapedSurfaces,CapekGustafssonSchab_MinimizationOfAntennaQualityFactor} is desired. The radiation Q-factor can be written as
\begin{equation}
\label{eq:Q1}
    Q = \frac{2\omega W + \left| P_\T{react} \right|}{2 P_\T{rad}},
\end{equation}
where
\begin{equation} \label{eq:storedEnergy}
    W = W_\T{e} + W_\T{m} = \frac{1}{4} \Ivec^\herm \M{W} \Ivec,
\end{equation}
is the cycle mean energy stored in electric and magnetic fields~\cite{Harrington_AntennaExcitationForMaximumGain,2018_Schab_Wsto,Gustaffson_QdisperssiveMedia_arXiv,CapekGustafssonSchab_MinimizationOfAntennaQualityFactor}
\begin{align}
    W_\T{e} &= \frac{1}{4\omega} \Ivec^\herm \XEmat \Ivec, \\
    W_\T{m} &= \frac{1}{4\omega} \Ivec^\herm \XMmat \Ivec,
\end{align}
where
\begin{equation}
    P_\T{rad} = \frac{1}{2} \Ivec^\herm \Rmat \Ivec
\end{equation}
is the cycle mean radiated power~\cite{Harrington_AntennaExcitationForMaximumGain,Harrington_FieldComputationByMoM} and
\begin{equation}\label{eq:Preact}
    P_\T{react} = 2 \omega \left( W_\T{m} - W_\T{e} \right) = \frac{1}{2} \Ivec^\herm \M{X} \Ivec
\end{equation}
is the cycle mean reactive power~\cite{Harrington_AntennaExcitationForMaximumGain,Harrington_FieldComputationByMoM,Gustaffson_QdisperssiveMedia_arXiv}. The matrices used in~\eqref{eq:storedEnergy}--\eqref{eq:Preact} are explicitly defined in Appendix~\ref{ap:matricesELMAG}. Notice that~\eqref{eq:Q1} is equivalent to
\begin{equation}
\begin{aligned}
\label{eq:QselfRes}
    Q = & \frac{\omega W }{P_\T{rad}} \\
    \T{s.t.} \quad & P_\T{react} = 0
\end{aligned}
\end{equation}
where the constraint~$P_\T{react} = 0$ enforces resonance. This last formulation is the starting point for formulating a lower bound on the Q-factor in the form of a \ac{QCQP} as
\begin{equation}
\begin{aligned} 
\label{eq:Q:QCQP}
\min \limits_\Ivec \quad & - \Ivec^\herm \Rmat \Ivec \\
    \T{s.t.} \quad & \Ivec^\herm \omega \Wmat \Ivec - 1 = 0 \\
    & \Ivec^\herm \Xmat \Ivec = 0.
\end{aligned}
\end{equation}
That constrained minimization~\eqref{eq:Q:QCQP} is equivalent to the minimization of~$Q$ in~\eqref{eq:QselfRes}, is guaranteed by the normative properties of cycle mean stored energy and its matrix~$\M{W}$, which is positive definite. If stored energy is a positive constant, then maximization of radiated power results in the minimization of the Q-factor. Vector~$\Ivec$ solving this optimization problem generates the lower bound on the radiation Q-factor, the value of which is given by~\eqref{eq:Q1} or~\eqref{eq:QselfRes}. With respect to a particular form of~\eqref{eq:Q:QCQP}, it is worth mentioning that a change~$\M{R_0} \to - \omega \M{W}, \omega \M{W} \to \M{R}_0$ is possible and this was used in~\cite{JelinekCapek_OptimalCurrentsOnArbitrarilyShapedSurfaces,CapekGustafssonSchab_MinimizationOfAntennaQualityFactor}. By means of numerical implementation, the form~\eqref{eq:Q:QCQP} is, nevertheless, more suitable since matrix~$\M{R}_0$ is not of full rank. 

\begin{figure}[!htb]
    \centering
    \includegraphics[width=\linewidth]{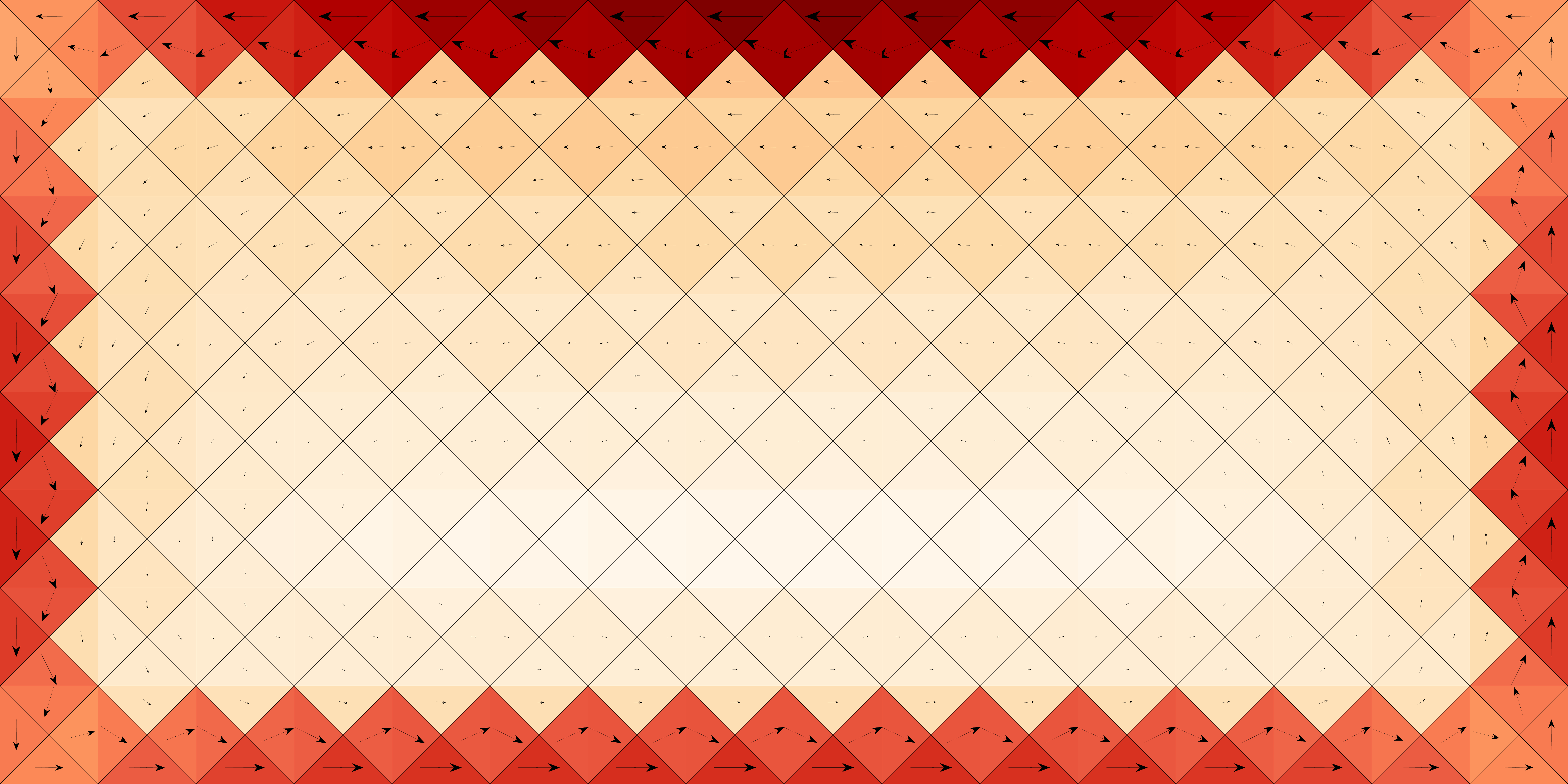}
    \caption{Current density corresponding to a lower bound on the Q-factor with self-resonant constraint. A rectangle with an aspect ratio of 2:1 is chosen as a support for optimal current density.}
    \label{fig:Qlb}
\end{figure}

For a numerical example, and as with ~\cite{JelinekCapek_OptimalCurrentsOnArbitrarilyShapedSurfaces,CapekGustafssonSchab_MinimizationOfAntennaQualityFactor}, a rectangle with an aspect ratio of 2:1 is chosen as a support for the optimal current density described by vector~$\M{I}$. The electrical size is set to~$ka = 0.5$, where~$k$ is the wavenumber in vacuum and~$a$ is the radius of the smallest sphere circumscribing the optimized region. \ac{RWG} functions defined over triangular mesh~\cite{RaoWiltonGlisson_ElectromagneticScatteringBySurfacesOfArbitraryShape}, see Fig.~\ref{fig:Qlb}, are used as basis functions in~\eqref{eq:currDenExpn}. 

The optimal current density, representing the minimum radiation Q-factor with self-resonant constraint is depicted in Fig.~\ref{fig:Qlb}.
The underlying \MATLAB implementation is briefly described in appendix~\ref{ap:exImplementation}. The corresponding normalized value of the radiation Q-factor is
\begin{equation} \label{eq:Qlb}
    (ka)^3 Q_\T{lb} = 4.6,
\end{equation}
where the normalization with $(ka)^3$ approximately removes the dependence on electrical size~$ka$ provided~\cite{GustafssonTayliEhrenborgEtAl_AntennaCurrentOptimizationUsingMatlabAndCVX,JelinekCapek_OptimalCurrentsOnArbitrarilyShapedSurfaces,CapekGustafssonSchab_MinimizationOfAntennaQualityFactor,Capek_etal_2019_OptimalPlanarElectricDipoleAntennas} that~$ka < 0.5$. The numerical implementation used fifth-order quadrature to evaluate \ac{MoM} reaction integrals. At the electric size of evaluation, the numerical value~\eqref{eq:Qlb} does not change when higher order quadrature and/or denser mesh is used.
\subsection{A Trade-Off Between Antenna Directivity, Dissipation Factor and Q-factor}
As a follow-up to the previous example about single-objective optimization and the lower bound on the radiation Q-factor, multi-objective optimization concerning minimum Q-factor, minimum dissipation factor and maximum directivity with the constraint on self-resonance is considered here.

\subsubsection{Minimum Dissipation Factor with Self-Resonant Constraint}\label{sec:minDelta}
Analogous to~\eqref{eq:QselfRes}, the problem of minimal self-resonant dissipation factor~\cite{Jelinek+etal2018,GustafssonCapekSchab_TradeOffBetweenAntennaEfficiencyAndQfactor} reads
\begin{equation}
\begin{aligned} \label{eq:delta}
    \min \limits_\Ivec \quad & \frac{P_\T{lost}}{P_\T{rad}} \\
    \T{s.t.} \quad & P_\T{react} = 0
\end{aligned}
\end{equation}
where
\begin{equation}
    P_\T{lost} = \frac{1}{2} \Ivec^\herm \RmatL \Ivec
\end{equation}
is the cycle mean lost power~\cite{GustafssonCapekSchab_TradeOffBetweenAntennaEfficiencyAndQfactor, JelinekCapek_OptimalCurrentsOnArbitrarilyShapedSurfaces, Jelinek+etal2018}, with matrix~$\M{R}_\rho$ detailed in Appendix~\ref{ap:matricesELMAG}. The \ac{QCQP} equivalent to~\eqref{eq:delta} is
\begin{equation}
\begin{aligned}
\label{eq:deltaQCQP}
    \min \limits_\Ivec \quad & - \Ivec^\herm \Rmat \Ivec \\
    \T{s.t.} \quad & \Ivec^\herm \RmatL \Ivec - 1 = 0 \\
    & \Ivec^\herm \Xmat_0 \Ivec = 0.
\end{aligned}
\end{equation}

\begin{figure}[!htb]
    \centering
    \includegraphics[width=\linewidth]{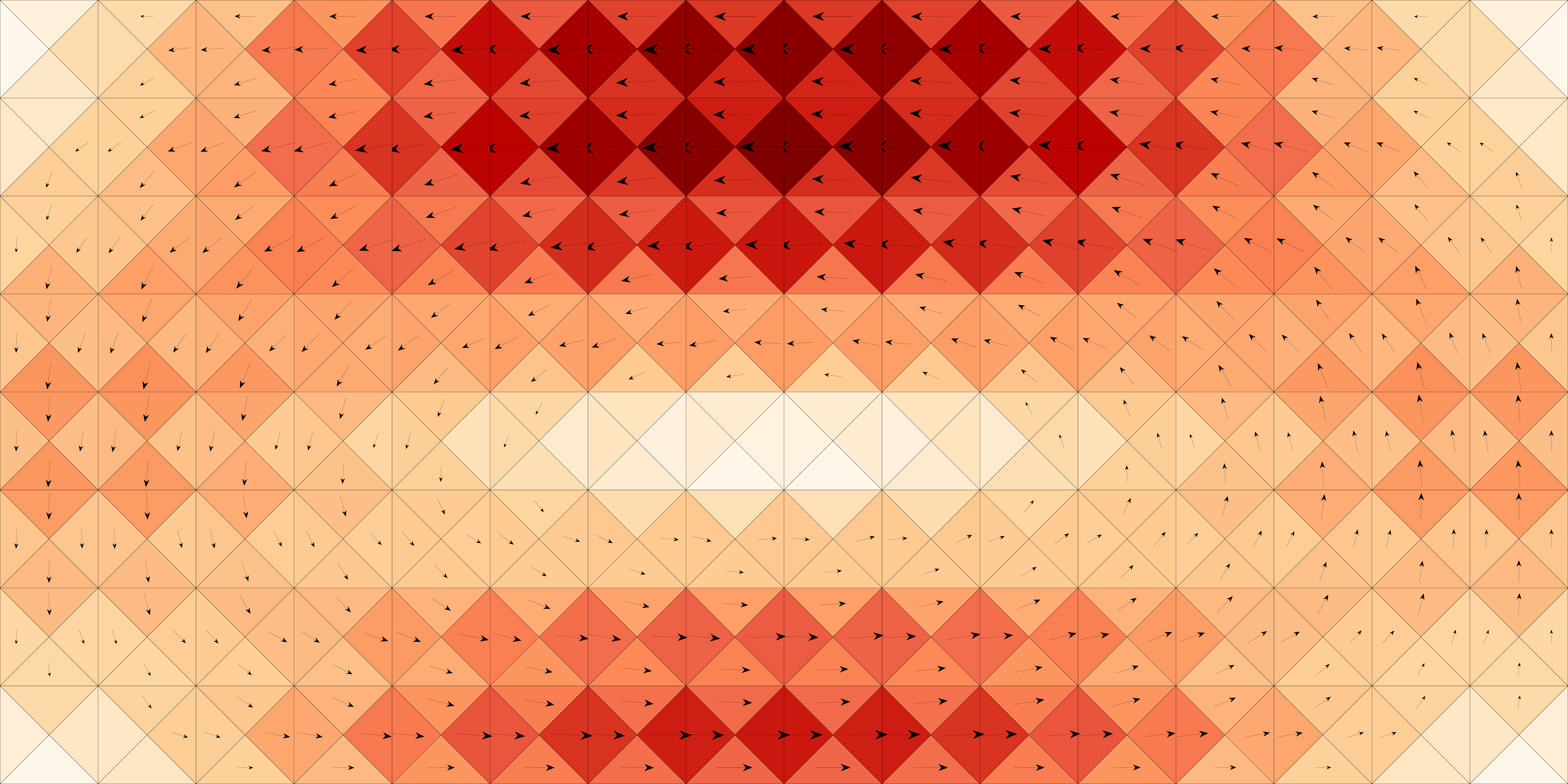}
    \caption{Current density representing the lower bound on dissipation factor  with self-resonant constraint. A rectangle with an aspect ratio of 2:1 is chosen as a support for optimal current density.}
    \label{fig:delta}
\end{figure}

The \MATLAB implementation of~\eqref{eq:deltaQCQP} is described in Appendix~\ref{ap:exImplementation} and, as with the same setup as in Section~\ref{sec:minQ}, the result is the normalized minimal self-resonant dissipation factor
\begin{equation} \label{eq:DELTAsr}
\frac{Z_0}{Z_\T{s}} (ka)^4 \delta_\T{lb} \approx 40,
\end{equation}
where the applied normalization removes~\cite{Jelinek+etal2018} the dependence on surface impedance~$Z_\T{s}$ and for~$ka < 0.5$ it also approximately removes the dependence on electrical size~$ka$. The free-space impedance~$Z_0$ is used to remove units. The optimal current density is depicted in Fig.~\ref{fig:delta}. Similar to the minimization of the Q-factor, here the optimal current density is also approximately composed of a mixture of electric-dipole like and magnetic-dipole like currents~\cite{CapekJelinek_OptimalCompositionOfModalCurrentsQ}.

\subsubsection{Pareto-Optimal Set of Dissipation and Q-factor with Self-Resonant Constraint}\label{sec:minQdelta}
\begin{figure}[!htb]
    \centering
    \includegraphics{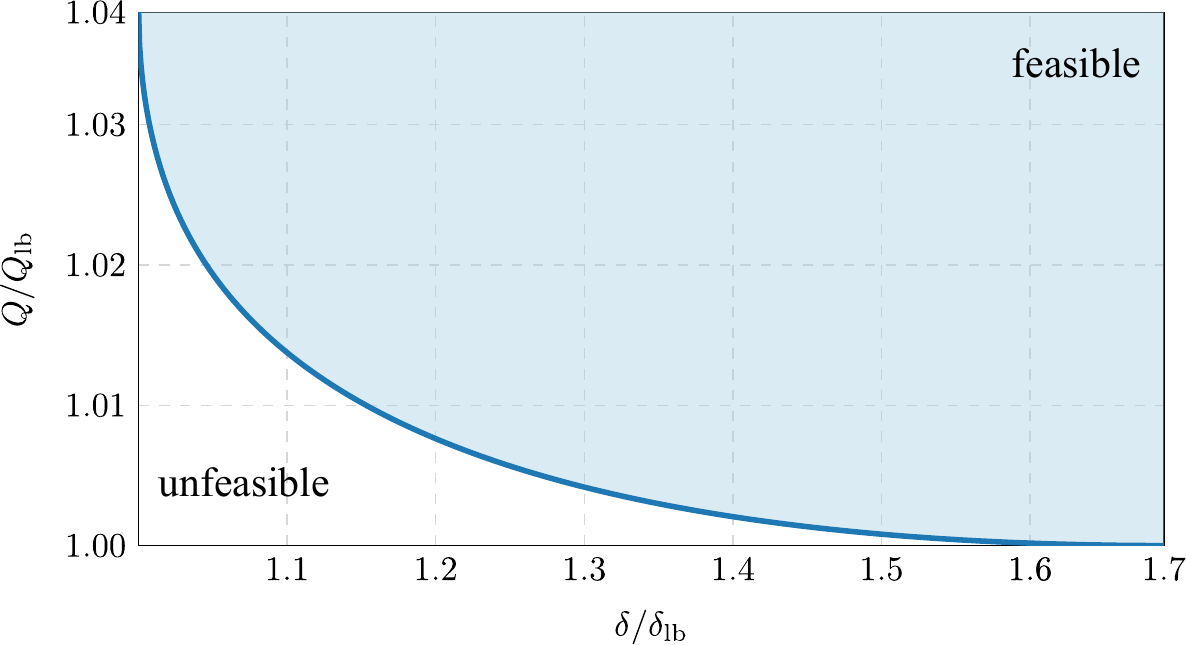}
    \caption{Pareto-optimal set of minimal radiation Q-factor and minimal dissipation factor with self-resonance constraint. Values are normalized by fundamental bounds on these metrics. The filled region is feasible, while the white region cannot be reached by any current distributed on the rectangle with aspect ratio~2:1.}
    \label{fig:paretoQdelta}
\end{figure}

Simultaneous minimization of Q-factor and self-resonant dissipation factor~\cite{GustafssonCapekSchab_TradeOffBetweenAntennaEfficiencyAndQfactor} results in a Pareto-optimal set according to Appendix~\ref{ap:pareto}. The optimization problem is
\begin{equation}
\begin{aligned} \label{eq:Qdelta}
    \min \limits_\Ivec \quad & (1 - c)Q + c\delta, \quad \forall c \in (0,1) \\
    \T{s.t.} \quad & P_\T{react} = 0,
\end{aligned}
\end{equation}
with the extreme cases $c=0$ and $c=1$ corresponding to the minimal Q-factor~\eqref{eq:QselfRes} and minimal dissipation factor~\eqref{eq:delta}, respectively. The equivalent \acs{QCQP} reads
\begin{equation}
\begin{aligned}
\label{eq:minQdelta}
    \min \limits_\Ivec \quad & - \Ivec^\herm \Rmat \Ivec \\
    \T{s.t.} \quad & \Ivec^\herm \left[(1 - c)\omega\Wmat + c \RmatL \right] \Ivec - 1 = 0, \quad \forall c \in (0,1) \\
    & \Ivec^\herm \Xmat_0 \Ivec = 0
\end{aligned}
\end{equation}
which is implemented according to Appendix~\ref{ap:exImplementation}. The resulting Pareto-optimal set is depicted in Fig.~\ref{fig:paretoQdelta} for the same setup as in previous examples and is seen to be of minimal extent since the two optimized parameters are almost nonconflicting. The current representing the minimal self-resonant dissipation factor is a practical and acceptable solution to the task.

\subsubsection{Pareto-Optimal Set of Dissipation Factor, Q-factor, and Directivity with Self-Resonant Constraint}\label{sec:minQdeltaD}
The Pareto-optimal set of Q-factor, dissipation factor, and directivity is obtained by the minimization of the convex combination of dissipation and Q-factor, as in the previous case, and by directivity taken as a constraint,
\begin{equation}
\begin{aligned} \label{eq:QdeltaD}
    \min \limits_\Ivec \quad & (1 - c)Q + c\delta, \quad \forall c \in (0,1) \\
    \T{s.t.} \quad & P_\T{react} = 0 \\
    & D = D_\T{c}, \quad \forall D_\T{c} \in (D_0(c),\infty),
\end{aligned}
\end{equation}
where it is noted that directivity is an unbounded metric that can reach arbitrarily high values~\cite{Bloch_PIEE1953} and that value~$D_0$ is the minimum directivity within the~$Q-\delta$ Pareto-optimal set from Fig.~\ref{fig:paretoQdelta}. For low values of parameter~$D_0$ the procedure must further omit all results that are not Pareto-optimal in the sense of maximal directivity.

The \acs{QCQP} equivalent to~\eqref{eq:QdeltaD} reads
\begin{equation}
\begin{aligned}
\label{eq:minQdeltaD}
    \min \limits_\Ivec \quad & - \Ivec^\herm \Rmat \Ivec \\
    \T{s.t.} \quad & \Ivec^\herm \left[(1 - c)\omega\Wmat + c \RmatL \right] \Ivec - 1 = 0, \quad \forall c \in (0,1) \\
    & \Ivec^\herm \Xmat \Ivec = 0 \\
    & \Ivec^\herm \left(8 \pi \Umat - D_\T{c} \Rmat \right) \Ivec = 0, \quad \forall D_\T{c} \in (D_0(c),\infty)
\end{aligned}
\end{equation}
where matrix~$\M{U}$ represents radiation intensity and is detailed in Appendix~\ref{ap:matricesELMAG}.

\begin{figure}[!htb]
    \centering
    \includegraphics{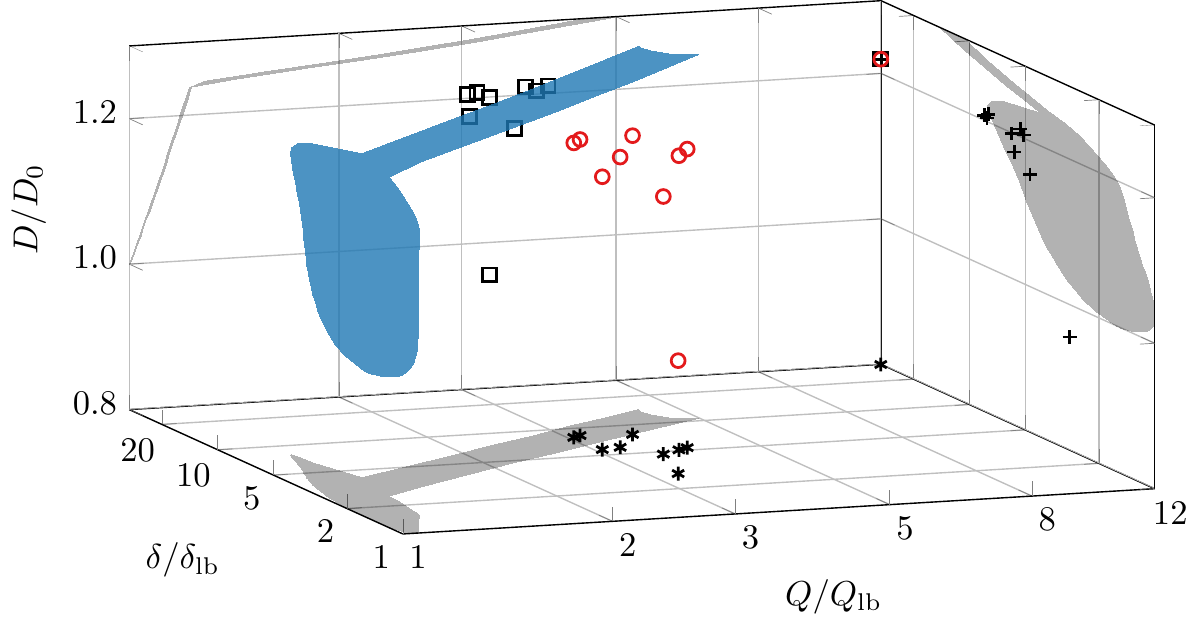}
    \caption{Pareto frontier of a broadside direction with $Q_\T{lb} = 37$, $\delta_\T{lb} = 0.017$ and $D_0 = 1.28$. Shadows denote its two-dimensional projections. The circle markers are the result of topology optimization, while the cross, square and asterisk markers are the corresponding projections. The projections of the Pareto frontier are not Pareto optimal in their entirety.}
    \label{fig:paretoQdeltaDbroadside}
\end{figure}

\begin{figure}[!htb]
    \centering
    \includegraphics{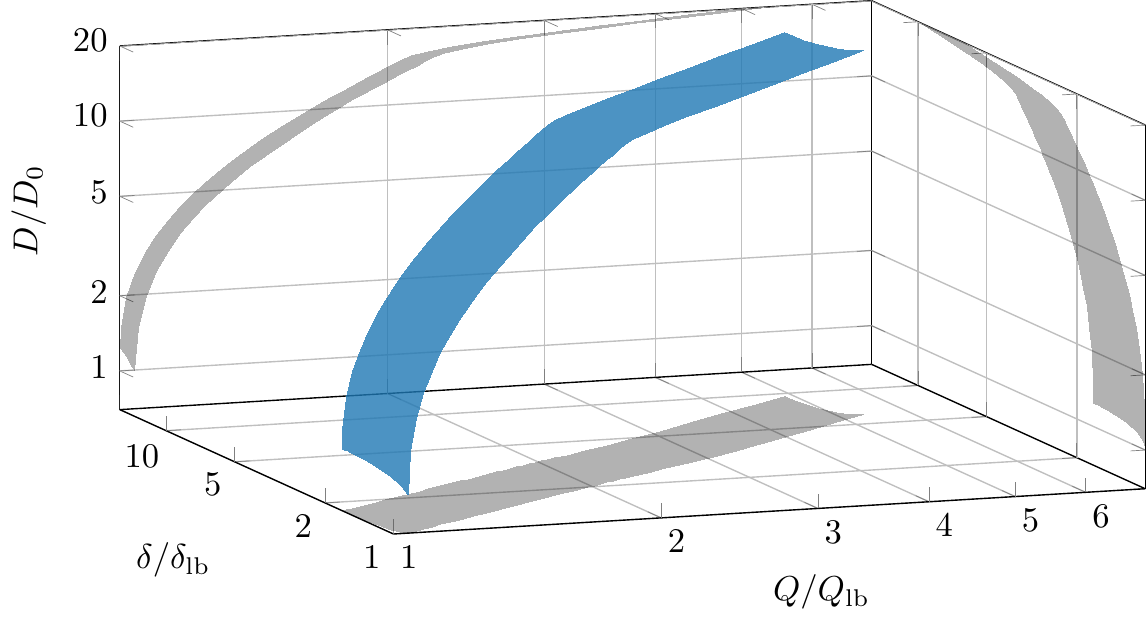}
    \caption{Pareto frontier of an end-fire direction with $Q_\T{lb} = 37$, $\delta_\T{lb} = 0.017$ and $D_0 = 0.19$. Plot follows the same scheme as in Fig.~\ref{fig:paretoQdeltaDbroadside}.}
    \label{fig:paretoQdeltaDendfire}
\end{figure}

Pareto frontiers corresponding to a rectangular patch treated in the previous examples are shown in Fig.~\ref{fig:paretoQdeltaDbroadside} and Fig.~\ref{fig:paretoQdeltaDendfire}. Figures also contain data coming from topology optimization~\cite{Capeketal_InversionFreeEvaluationOfNearestNeighborsInMoM,Capeketal_ShapeSynthesisBasedOnTopologySensitivity} over the same patch fed in the middle at the top, see Fig.~\ref{fig:delta}, by a delta-gap source. These later data show the feasibility of the fundamental bound.

The MATLAB implementation of the optimization problem is described in  Appendix~\ref{impl:QdeltaD}.

\subsection{Substructure Bounds}\label{sec:affine}
When basis functions~$\V{\psi}_n$ are sufficiently localized, an interesting variation in the formulation of fundamental bounds is to optimize only a part of current vector~$\M{I}$ (substructure consisting of only selected basis functions) lying within a \emph{controllable} part of the structure~\cite{Gustafsson_OptimalAntennaCurrentsForQsuperdirectivityAndRP, 2021_Jelinek_OPEX}, leaving the rest of the structure (rest of the current vector) to evolve according to Maxwell's equations (this part is called \emph{uncontrollable}). Under the \ac{MoM} formulation~\cite{Harrington_FieldComputationByMoM} of field integral equation~$\M{Z} \M{I} = \M{V}$, see also Appendix~\ref{ap:matricesELMAG}, the electromagnetic description of the system can be partitioned as
\begin{equation}
\label{eq:cu1}
\mqty [
\M{Z}_\T{cc} & \M{Z}_\T{cu}\\
\M{Z}_\T{uc} & \M{Z}_\T{uu}
]
\mqty[\M{I}_\T{c}\\ \M{I}_\T{u}] 
= 
\mqty[ \M{V}_\T{c}\\ \M{V}_\T{u} ]
\end{equation}
where index ``c'' denotes ``controllable'', ``u'' denotes ``uncontrollable'',~$\M{Z}$ is the system matrix~\cite{Harrington_FieldComputationByMoM} and~$\M{V}$ represents excitation~\cite{Harrington_FieldComputationByMoM}. 

The partitioning~\eqref{eq:cu1} offers the possibility to solely control the current in the controllable region described by vector~$\M{I}_\T{c}$. To that point, the uncontrollable current~$\M{I}_\T{u}$ is eliminated from~\eqref{eq:cu1} using 
\begin{equation}
\mqty [
\M{Z}_\T{uc} & \M{Z}_\T{uu}
]
\M{I}
= 
\M{V}_\T{u}
\label{eq:cuConst}
\end{equation}
as an affine constraint. According to Appendix~\ref{ap:affineConstraints}, this constraint is equivalent to an affine transformation of variables~$\M{I} \to \M{x}$
\begin{equation}
    \label{eq:cu2}
\M{I} = \M{t} + \M{T} \M{x},
\end{equation}
which can be used to remove this constraint from the optimization problem. This choice of new basis vectors (columns of matrix~$\M{T}$) guarantee that even when total current~$\M{I}$ does not satisfy~\eqref{eq:cu1}, the current in the uncontrollable region~$\M{I}_\T{u}$ is strictly governed by Maxwell's equations, i.e., by~\eqref{eq:cuConst}, which reflects not only the reaction of the uncontrollable region on incident wave~$\M{V}_\T{u}$, but also the reaction on the field~$-\M{Z}_\T{uc} \M{I}_\T{c}$ generated by the controllable region. Such a procedure is analogous to the construction of Green's function in the presence of scattering objects or boundaries~\cite{Tai_Dyadic_green_functions}.

\begin{figure}[htb!]
     \centering
     \begin{subfigure}[b]{0.45\textwidth}
         \centering
         \includegraphics[width=\textwidth]{figures/mesh_maxAbsorption.pdf}
     \end{subfigure}
     \hfill
     \begin{subfigure}[b]{0.45\textwidth}
         \centering
         \includegraphics[width=\textwidth]{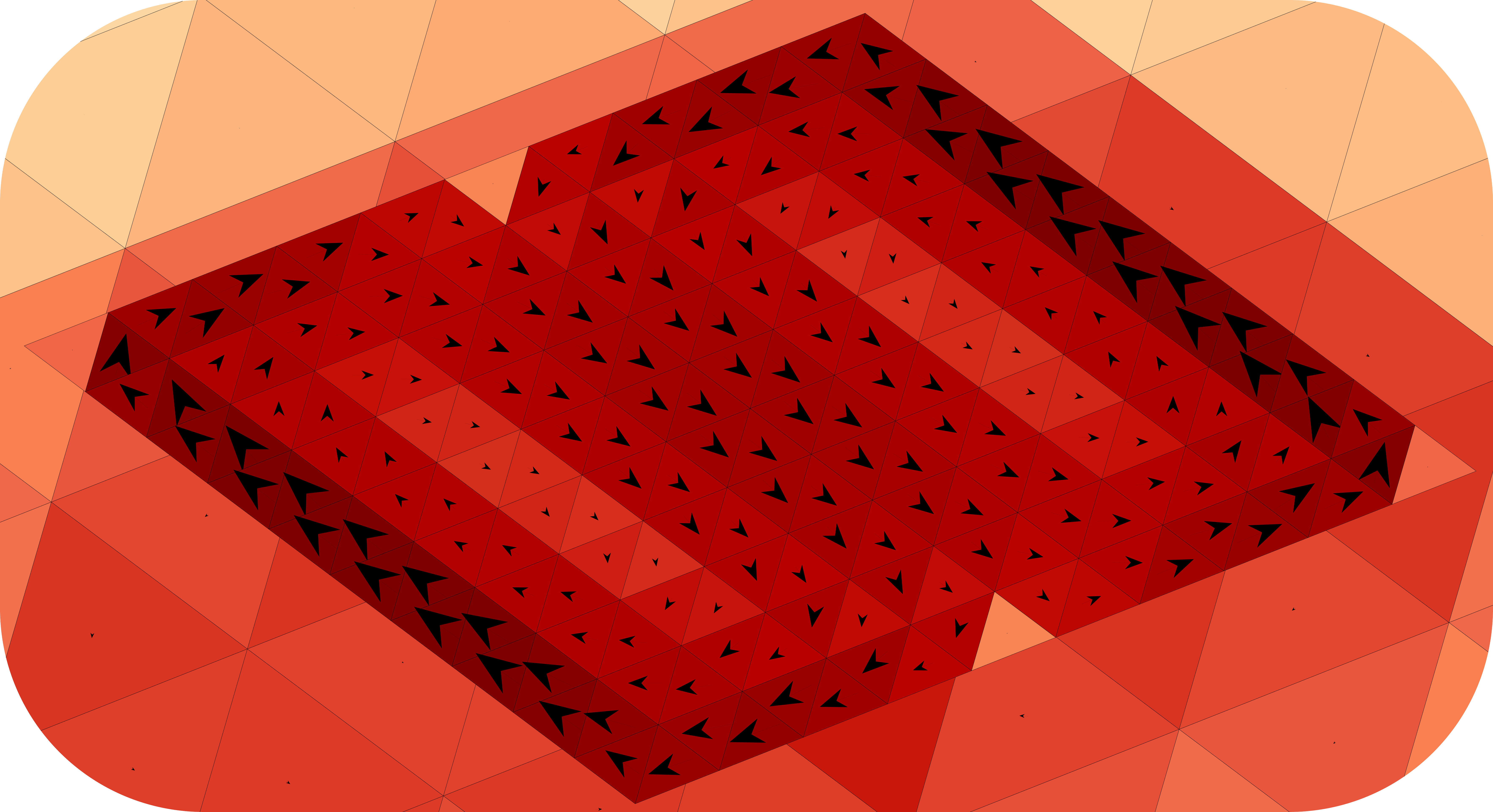}
     \end{subfigure}
        \caption{(left) Triangular mesh of the current supporting region. The blue color represents the controllable region, while the yellow region shows an uncontrollable region. The ratio of edge lengths of the yellow and blue regions is approximately 3:1. The controllable patch is placed at a height of ~$W/6$ over the uncontrollable ground plane, $W$ being the edge length of the controllable patch. (right) Optimal current density realizing maximum total absorption and zero back-scattering of a plane wave impinging normally on the structure from top to bottom.}
        \label{fig:mesh_maxAbsorp}
\end{figure}

As an example of this procedure, suppose a lossy uncontrollable region (the yellow square patch in the left panel of Fig.~\ref{fig:mesh_maxAbsorp}) acting as a finite-sized ground plane and a lossy controllable region (the blue square patch of edge length~$W$ in the left panel of Fig.~\ref{fig:mesh_maxAbsorp} fully controlled by optimization) the function of which is to provide the highest absorption in the entire structure and, simultaneously, vanishing back-scattering of a plane wave impinging perpendicularly on the structure and being polarized along the edge of the controllable region.

The resulting \ac{QCQP} reads
\begin{equation}
\begin{aligned}
	\min \limits_{\M{x}} \quad & - \M{I}^\herm \M{R}_{\rho} \M{I} \\
	\T{s.t.} \quad &\M{I}^\herm \M{Z} \M{I} = \M{I}^\herm \M{V} \\
	&\M{I} = \M{t} + \M{T} \M{x},
\end{aligned}  
\label{eq:Pa_opt}
\end{equation}
where the first constraint enforces the conservation of complex power~\cite{2020_Gustafsson_NJP}, being a relaxed version of full system equation~$\M{Z} \M{I} = \M{V}$ and actually representing two real quadratic constraints. The second constraint aggregates all affine constraints imposed on the problem, i.e., division into a controllable/uncontrollable region~\eqref{eq:cuConst} and the constraint realizing zero back-scattering which is given by equation
\begin{equation}
    \mqty[\M{F}_\theta \\ \M{F}_\phi] \M{I} = \M{0},
\end{equation}
where far-field vectors~$\M{F}$ are detailed in Appendix~\ref{ap:matricesELMAG}.

\begin{figure}[htb!]
     \centering
     \begin{subfigure}[b]{0.6\textwidth}
         \centering
         \includegraphics[width=\textwidth]{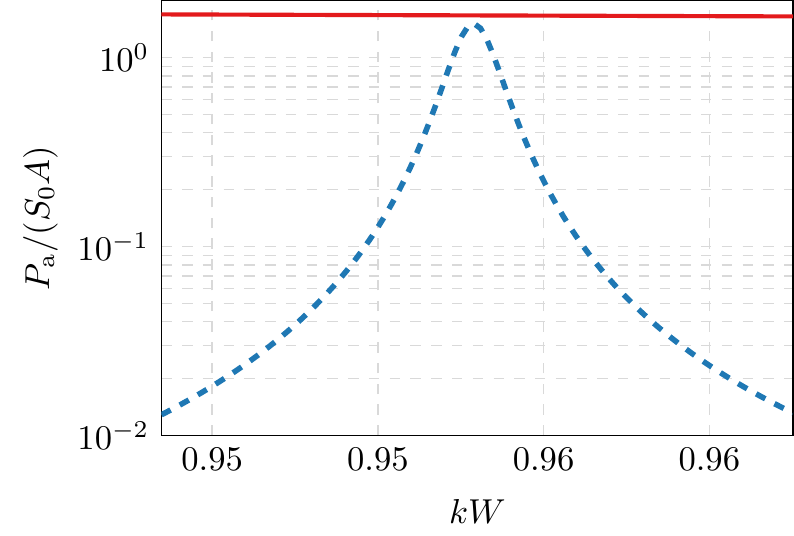}
     \end{subfigure}
     \hfill
      \begin{subfigure}[b]{0.35\textwidth}
         \centering
         \includegraphics[width=\textwidth]{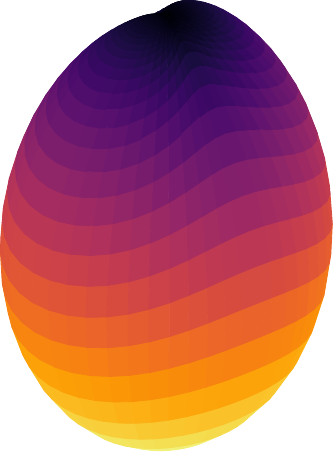}
     \end{subfigure}
    \caption{(left) Frequency sweep of the normalized optimal absorption (red solid curve) and absorption realized by an electric-coupled resonator~\cite{2006_Schurig_APL} over the same ground plane (blue dashed curve). The incident power flux is denoted by~$S_0$ and $A$ abbreviates the geometric cross-section of the structure. (right) Scattering diagram corresponding to the optimal current density from Fig.~\ref{fig:mesh_maxAbsorp} for a plane wave impinging from top to bottom.}
\label{fig:maxAbsorp}
\end{figure}

The results for the considered setup are depicted in Fig.~\ref{fig:maxAbsorp}, showing a frequency sweep of the normalized optimal absorption and a representative scattering diagram at the central frequency, which clearly presents the desired vanishing backscattering. The optimal current density at the central frequency is depicted in the right panel of Fig.~\ref{fig:mesh_maxAbsorp} and suggests an electric-coupled resonator proposed in~\cite{2006_Schurig_APL} as a potential design that can fulfill the optimization requirements. That this is nearly the case is shown by the realized absorption (see the dashed curve in Fig.~\ref{fig:maxAbsorp}) of this resonator placed over the considered ground plane. The considered surface impedance is~$Z_\T{s} = 0.01 \, \Omega$.

The \MATLAB implementation of the optimization problem is described in  Appendix~\ref{impl:SubStructure}.
\section{Numerical Precision and Computational Efficiency}
Numerical precision and the computational complexity of fundamental bounds, mostly depending on the discretization of the support of the optimal current density, are considered in this section. The quality factor is used as the optimized metric. 

\begin{figure}[htb!]
    \centering
    \includegraphics{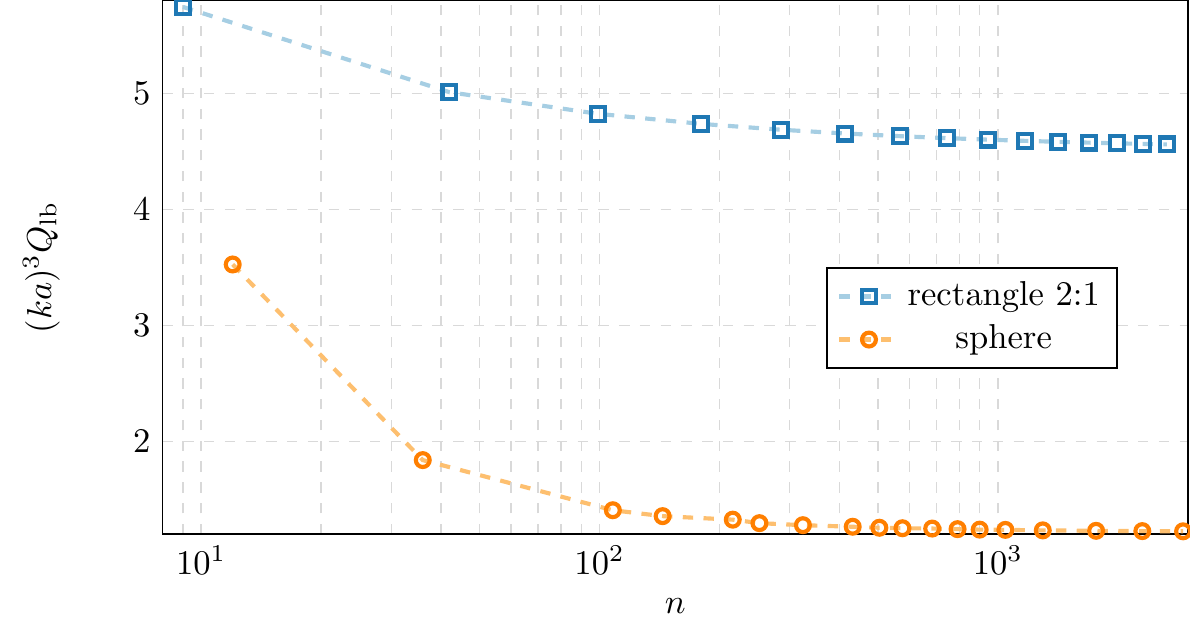}
    \caption{Normalized lower bound on Q-factor for a varying number of dicretization elements $n$. The sphere and rectangle of edge length ratio 1:2 are considered. A quadrature of the fifth order is used in all cases to evaluate reaction integrals resulting from the MoM description.}
    \label{fig:minQprecision}
\end{figure}

\begin{figure}[htb!]
    \centering
    \includegraphics{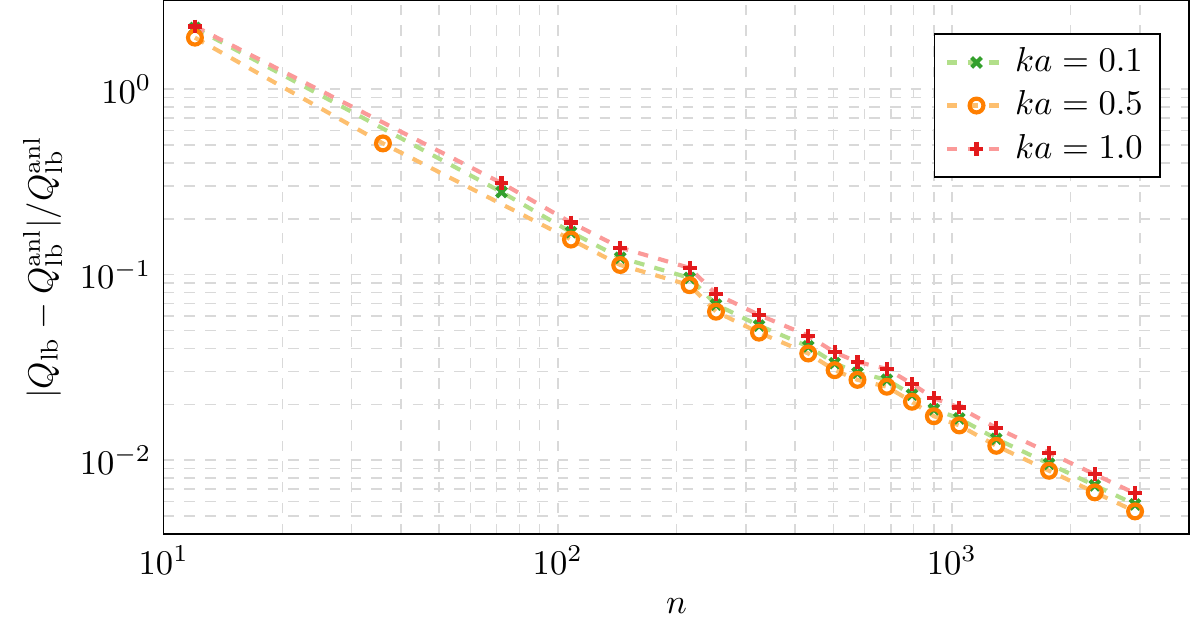}
    \caption{Relative error of the lower bound to Q-factor for a varying number of dicretization elements $n$ on a sphere. The analytically known value of the minimum quality factor~\cite{CapekJelinekHazdraEichler_QofSphere} is denoted as~$Q^{\T{anl}}_\T{lb}$.}
    \label{fig:minQerrSphr}
\end{figure}

Numerical precision is addressed in Fig.~\ref{fig:minQprecision} and Fig.~\ref{fig:minQerrSphr}. Using the analytically known value of the minimum quality factor of a spherical shell~\cite{CapekJelinekHazdraEichler_QofSphere}, it is observed that the achieved gain in numerical precision is approximately one digit per one order in the number of discretization elements as in the case of self-resonant dissipation factor, see~\cite[App.~C]{Jelinek+etal2018}.

\begin{figure}[htb!]
    \centering
    \includegraphics{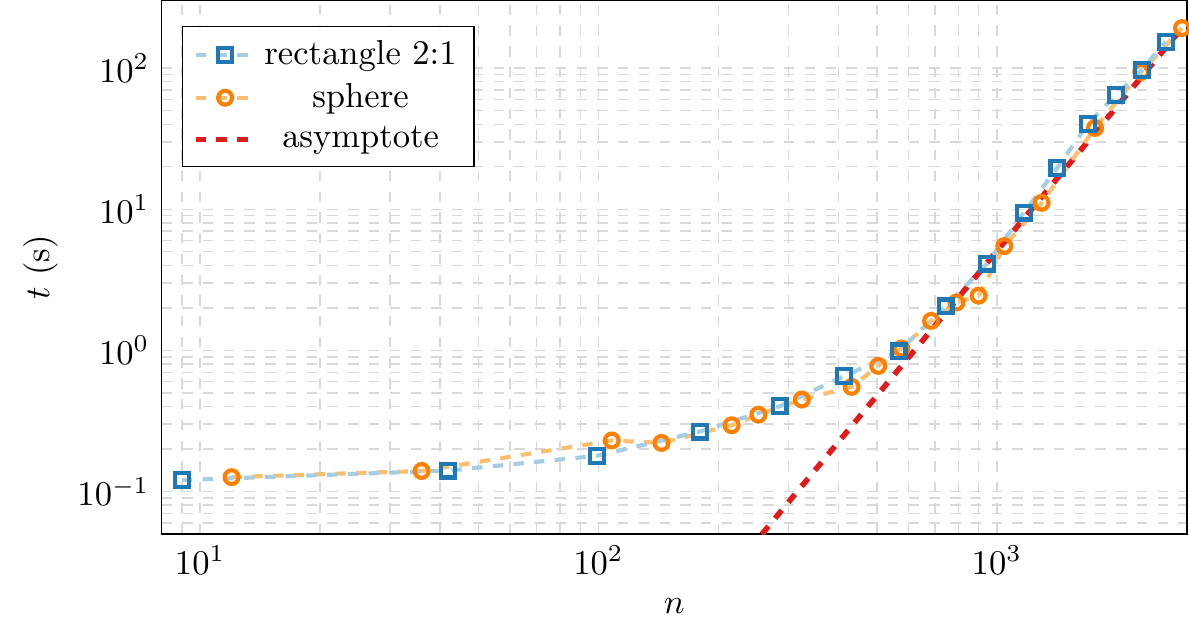}
    \caption{Computational cost of solution to the optimization problem~\eqref{eq:Q:QCQP}  (the evaluation of necessary matrices is not counted) as a function of number of dicretization elements $n$. The same setup as~Fig.~\ref{fig:minQprecision} is considered. A mobile Intel processor Intel(R) Core(TM) i7-8850H CPU @ 2.60GHz was employed. The figure also shows an asymptote with~$n^{3.4}$ slope.}
    \label{fig:minQtime}
\end{figure}

Numerical precision is related to computational efficiency and depends on the number of discretization elements. The dependence is shown in Fig.~\ref{fig:minQtime}.
 
The results show that the algorithmic complexity of evaluating fundamental bounds is slightly higher than the solution of a linear equation system as implemented in MATLAB via the \mtbtext{mldivide} function. For small sizes of the underlying problem, the computation cost is dominated by an overhead of underlying functions with a low scaling factor. For larger sizes, $n > 500$, the computational cost is dominated by the evaluation of eigenvalues with an approximate scaling of~$n^3$.

Both solvers used in this work, and described in Appendix~\ref{ap:coreFunctions}, are iterative, using standard methods of convex optimization such as Newton's or the simplex method. To increase computational efficiency, it is therefore advantageous to begin with a low discretization which gives approximate values of Lagrange's multipliers~$\mu$, see Appendix~\ref{ap:QCQP}. Those are then used as the initial prediction for higher discretization to save the number of iteration steps.

\section{AToM Package: Fundamental Bounds}

Examples in Section~\ref{sec:examples} were all evaluated using the ``FunBo'' package attached to the \acf{AToM}~\cite{atom}, which is a numerical tool for the analysis and synthesis of electromagnetic structures developed at the Department of Electromagnetic Field at FEE CTU in Prague.

The package, including scripts for the above studied examples can be found on \ac{AToM}'s web page \href{http://antennatoolbox.com/fundamentalBounds}{antennatoolbox.com/fundamentalBounds}. As a supplement to the commentaries found in the scripts,  Appendix~\ref{ap:exImplementation} tries to briefly explain the most important parts of the underlying codes. The procedure to choose the right solver for a given problem is then shown in Appendix~\ref{ap:flowChart}.

For the purpose of this text, the matrices shown in Appendix~\ref{ap:matricesELMAG} and the entering of optimizations were evaluated in the \acf{AToM}, which, therefore, must be included in the \MATLAB path for examples to run properly. The package for fundamental bounds can, however, also be used with matrices supplied by the user. Notably, the core functions introduced in Appendix~\ref{ap:coreFunctions} can be applied to arbitrary \ac{QCQP} satisfying the conditions listed in Appendix~\ref{ap:QCQP}.

The contents of the package includes:
\begin{itemize}[align=left]
    \item[\mtbtext{+controllableRegion}] Functions in this name space treat affine constraints and controllable/uncontrollable region problems.
    \item[\mtbtext{+optimProblems}] Generic functions computing fundamental bounds on frequently used antenna and scattering metrics are prepared in this name space.
    \item[\mtbtext{+solvers}] The solvers to \ac{QCQP} are contained in this subspace. All solvers use dual formulation.
    \item[\mtbtext{examples}] Folder with examples, including those detailed in Section~\ref{sec:examples}.
\end{itemize}
\section*{Acknowledgement}
This work has been supported by the Czech Science Foundation under project~\mbox{No.~19-06049S} and by the Grant Agency of the Czech Technical University in Prague under project \mbox{No.~SGS19/168/OHK3/3T/13}.

\appendix
\appendixpage

\section{Matrix Representation} \label{ap:matricesELMAG}
Method of moments~\cite{Harrington_FieldComputationByMoM,Gibson_MoMinElectromagnetics,ChewTongHu_IntegralEquationMethodsForElectromagneticAndElasticWaves} is a numerical tool which converts a linear integro-differential operator equation to a system of linear equations. Typically, a symmetric complex matrix represents an operator, an unknown is represented by the vector of expansion coefficients~$\M{I}$ and excitation is commonly denoted by vector~$\M{V}$. The representation employs a reaction  product\footnote{If surface structures are treated, the volume integral reduces to a surface integral.}~\cite{Rumsey_ReactionConceptInElectromagneticTheory}
\begin{equation}
    \langle \bm{f}, \bm{g}\rangle = \int\limits_{V_\T{s}} \bm{f} (\bm{r)} \cdot \bm{g} (\bm{r}) \ \T{d}V,
\end{equation}
where $\bm{f},\bm{g}$ are scalar or vector functions. Within this text, the electromagnetic interaction is described by an \ac{EFIE}~\cite{Harrington_FieldComputationByMoM} with real-valued basis functions~$\V{\psi}$ and the system equation
\begin{equation}
    \left( \Zmat_\rho + \Zmat_0 \right) \Ivec = \Vvec,
\end{equation}
where the system matrix consists of material part~$\Zmat_\rho$ and vacuum part~$\Zmat_0$. The elements of the material matrix $\Zmat_\rho = \RmatL + \T{j} \Xmat_\rho$ are
\begin{equation}
    z^\rho_{mn} = \langle \basisFcn_m,\V{\rho} \basisFcn_n \rangle
\end{equation}
with 
\begin{equation}
    \V{\rho} =  -  \T{j} \dfrac{Z_0}{k} \V{\chi}^{-1}
\end{equation}
where~$\V{\chi}$ is the electric susceptibility tensor. In the case of a highly conducting obstacle, the term~$\V{\rho} \basisFcn_n$ is substituted by~$Z_\T{s} \basisFcn_n$ with~$Z_\T{s}$ being a surface impedance~\cite{Jackson_ClassicalElectrodynamics,SenoirVolakis_ApproximativeBoundaryConditionsInEM}.

Elements of matrix~$\Zmat_0 = \Rmat + \T{j} \Xmat_0$ are given by~\cite{Harrington_FieldComputationByMoM}
\begin{equation}\label{eq:impedance_matrix_element}
    z^0_{mn} = \T{j} Z_0 k \left( \langle \basisFcn_m, \langle G, \basisFcn_n \rangle \rangle - \frac{1}{k^2} \langle \nabla \cdot \basisFcn_m, \langle G, \nabla \cdot \basisFcn_n \rangle \rangle \right),
\end{equation}
where $G$ denotes the free-space Green's function
\begin{equation}
    G(\bm{r}, \bm{r}') = \dfrac{\expn{- \T{j}k|\bm{r}-\bm{r}'|}}{4 \pi |\bm{r}-\bm{r}'|}.
\end{equation}

Finally, the elements of excitation vector~$\M{V}$ is calculated as
\begin{equation}
    V_m = \langle \basisFcn_m, \V{E}^\T{i} \rangle
\end{equation}
with~$\V{E}^\T{i}$ being the incident electric field.

According to~\cite{Harrington_AntennaExcitationForMaximumGain,2018_Schab_Wsto,Gustaffson_QdisperssiveMedia_arXiv}, the matrix representing the cycle mean stored energy is given as
\begin{equation}
    \Wmat = \dfrac{\partial \Xmat_0}{\partial \omega},
    \label{eq:Wmat}
\end{equation}
where the derivative of the impedance matrix with respect to angular frequency $\omega$
\begin{multline}
    \omega \dfrac{\partial z^0_{mn}}{\partial \omega} = 
     \T{j} Z_0 k \left( \langle \basisFcn_m, \langle G, \basisFcn_n \rangle \rangle + \frac{1}{k^2} \langle \nabla \cdot \basisFcn_m, \langle G, \nabla \cdot \basisFcn_n \rangle \rangle \right) +
     \\
    +Z_0 k \left( \langle \basisFcn_m, \langle k |\bm{r}-\bm{r}'|G, \basisFcn_n \rangle \rangle - \dfrac{1}{k^2} \langle \nabla \cdot \basisFcn_m, \langle k |\bm{r}-\bm{r}'|G, \nabla' \cdot \basisFcn_n \rangle \rangle \right)
\end{multline}
is needed. 

The knowledge of matrices representing reactive power and stored energy can be used to derive matrices~\cite{Gustaffson_QdisperssiveMedia_arXiv,CapekGustafssonSchab_MinimizationOfAntennaQualityFactor} representing the cycle mean energy stored in the electric field
\begin{equation}
\XEmat = \dfrac{\omega \Wmat - \Xmat_0}{2}
\end{equation}
and in the magnetic field
\begin{equation}
    \XMmat = \dfrac{\omega \Wmat + \Xmat_0}{2}.
\end{equation}

In the presence of material media with no temporal dispersion, the substitution~$\M{X}_0 \to \M{X}_0 + \M{X}_\rho$ in~\eqref{eq:Wmat} also gives the correct values of stored energy. Temporally dispersive media must, in many cases, be treated differently~\cite{Gustaffson_QdisperssiveMedia_arXiv}.

Radiation intensity matrix~$\Umat$ is related to the matrix of electric far-field~$\M{F}$ projected into a given direction~$\hat{\V{e}}$, typically~$\bm{\theta}_0$ or~$\bm{\varphi}_0$~\cite{JelinekCapek_OptimalCurrentsOnArbitrarilyShapedSurfaces}. The connection between the two reads
\begin{equation}\label{eq:Umat}
    \Umat \left(\hat{\V{d}}, \hat{\V{e}}\right) = \dfrac{\Fmat^\herm \left(\hat{\V{d}}, \hat{\V{e}}\right) \Fmat \left(\hat{\V{d}}, \hat{\V{e}}\right)  }{2 Z_0},
\end{equation}
and the matrix elements of the projected electric far-field matrix reads
\begin{equation}
   F_n \left(\hat{\V{d}}, \hat{\V{e}}\right) = \dfrac{- \T{j} Z_0 k}{4 \pi} \langle \expn{\T{j} k \hat{\V{d}} \cdot \bm{r}}, \hat{\V{e}} \cdot \basisFcn_n \rangle,
\end{equation}
where $\hat{\V{d}}$ is the unit vector in the direction of observation.

\section{Quadratically Constrained Quadratic Program} \label{ap:QCQP}
The solution to a general \ac{QCQP} is, in this text, attempted by in-house routines contained in the MATLAB package ``FunBo'' attached to the~\acf{AToM}~\cite{atom}, a numerical tool developed at the department of electromagnetic field at CTU FEE in Prague. The following form of \ac{QCQP} is considered
\begin{align}
\label{eq:genProb}
        \min \limits_\Ivec \quad & \Ivec^\herm \M{A} \Ivec + \RE[\Ivec^\herm \V{a}] + \alpha \\
        \label{eq:genProbConstr}
    \T{s.t.} \quad & \Ivec^\herm \M{B}_i \M{I} + \RE[\Ivec^\herm \V{b}_i] + \beta_i = 0; \quad \forall i = 1, \ldots, m
\end{align}
where
\begin{align*}
& \M{I}, \V{a}, \V{b}_i \in \mathbb{C}^{n \times 1}, \\
& \M{A}, \M{B}_i \in \mathbb{C}^{n \times n}, \\
& \alpha, \beta_i \in \mathbb{R}, \\
& \M{B}_1 \succ 0,
\end{align*}
all matrices are Hermitian and a solution exists. Lagrange's function, associated with~\eqref{eq:genProb}--\eqref{eq:genProbConstr}, reads
\begin{multline} \label{eq:lgrFcn}
    L(\M{I},\mu_1,\ldots,\mu_m) = \Ivec^\herm
    \left(\M{A}-\sum \limits_{i=1}^m \mu_i \M{B}_i \right)
    \Ivec + \\ 
    + \RE \left[\Ivec^\herm
    \left(\V{a}-\sum \limits_{i=1}^m \mu_i \V{b}_i\right) \right]+ \alpha - \sum \limits_{i=1}^m \mu_i \beta_i,
\end{multline}
where~$\mu_i$ are Lagrange's multipliers.

The solution is approached via the dual formulation in which the dual function
\begin{equation}
    \label{eq:g:def}
    g \left(\mu_1, \ldots, \mu_m \right) = \inf  \limits_\Ivec L(\Ivec,\mu_1,\ldots,\mu_m)
\end{equation}
is constructed and later maximized over variables~$\mu_i$. This gives~\cite{BoydVandenberghe_ConvexOptimization} the lower bound\footnote{It might happen that the dual problem provides a solution which does not satisfy all constraints~\eqref{eq:genProbConstr}. In such a case the so-called dual gap~\cite{BoydVandenberghe_ConvexOptimization} appears and the solution to the dual problem can only be taken as a lower bound to the original (primal) problem.} to~\eqref{eq:genProb}--\eqref{eq:genProbConstr}.

Two cases must be considered. The first case governs the situation when
\begin{equation} \label{eq:linT}
    \V{a} = \V{0}, \quad \V{b}_i = \V{0}, \ \forall i \in \{ 1,2,\ldots,m \}, 
\end{equation}
in which the stationary points of Lagrange's function read
\begin{equation} \label{eq:gep}
    \left(\M{A}-\sum \limits_{i=2}^m \mu_i \M{B}_i \right)
    \Tilde{\Ivec} =
    \lambda \M{B}_1 \Tilde{\Ivec}
\end{equation}
and generate the dual function
\begin{equation} \label{eq:dual}
    g \left(\mu_1, \ldots, \mu_m \right) = \begin{cases}
    \alpha - \sum \limits_{i=1}^m \mu_i \beta_i; & \T{if} \ \mu_1 = \min \{ \lambda \}, \\
    \\
    - \infty; & \T{otherwise}.
    \end{cases}
\end{equation}
Minimizer~$\hat{\Ivec}$ for~\eqref{eq:genProb}--\eqref{eq:genProbConstr} is a linear combination of eigenvectors $\Tilde{\Ivec}$ associated with the eigenvalue~$\mu_1 = \min \{ \lambda \}$ evaluated at the maximum of the dual function~\eqref{eq:dual} and fulfilling all constraints~\eqref{eq:genProbConstr}.

In the second case, when at least one of the vectors~$\M{a}, \M{b}_i$ is nonzero, the stationary points of Lagrange's function read
\begin{equation}
\label{eq:Istat2}
    \dot{\Ivec} = -\frac{1}{2} \left(\M{A}-\sum \limits_{i=1}^m \mu_i \M{B}_i \right)^{-1} \left( \V{a} - \sum \limits_{i=1}^m \mu_i \V{b}_i \right),
\end{equation}
which is related to the dual function
\begin{equation} \label{eq:dualLin}
    g \left(\mu_1, \ldots, \mu_m \right) 
     = \begin{cases}
    \dfrac{1}{4} \dot{\Ivec}^\herm
    \left(\V{a} - \sum \limits_{i=1}^m \mu_i \V{b}_i \right)+ \alpha
    - \sum \limits_{i=1}^m \mu_i \beta_i; & \T{if} \ \mu_1 < \mu_1^\T{min}, \\
    \\
    - \infty; & \T{otherwise}.
    \end{cases}
\end{equation}
The condition $\mu_1 < \mu_1^\T{min} = \min \{ \lambda \}$ results from the demand of~\eqref{eq:Istat2} being a local minimum (the Hessian matrix of Lagrange's function being positive definite). In this case, minimizer~$\hat{\M{I}}$ for~\eqref{eq:genProb}--\eqref{eq:genProbConstr} is equivalent to $\dot{\M{I}}$ given by~\eqref{eq:Istat2} in the maximum of the dual function~\eqref{eq:dualLin}.

Due to its construction~\eqref{eq:g:def}, the dual function~$g \left(\mu_1, \ldots, \mu_m \right)$ is convex and its maximum can be found by standard tools, such as Newton's method or the simplex method. Within the aforementioned package, this solution is provided by MATLAB functions \mtbtext{QNCQPQuadLin()}, \mtbtext{minLinStB1()}, \mtbtext{minAstBn()}, and \mtbtext{minAstB1()}, which are briefly described in Appendix~\ref{ap:coreFunctions}. The appropriate function to solve a given \ac{QCQP} can be chosen according to Appendix~\ref{ap:flowChart}.

\section{Flow Chart} \label{ap:flowChart}
\begin{figure}[t!hb]
\centering
\begin{tikzpicture}[node distance=3.4cm, auto]
\tikzstyle{decision} = [diamond, draw, fill=blue!20, text width=5.5em, text badly centered, inner sep=0pt]
\tikzstyle{block} = [rectangle, draw, fill=blue!20, text width=8.5em, text centered, rounded corners, minimum height=3em]
\tikzstyle{line} = [draw, -latex']
\tikzstyle{cloud} = [draw, ellipse,fill=red!20, minimum height=1em]

\node [block] (definition) {Define Problem};
\node [block, below of=definition, yshift=1.5cm] (quadFun) {Establish Quadratic Functionals};
\node [cloud, left of=quadFun] (operators) {Operators};
\node [cloud, left of=definition] (metrics) {Metrics};
\node [block, right of=quadFun, xshift=0.5cm] (QCQP) {Define QCQP, where $\mathbf{B}_1 \succ 0$};
\node [decision, below of=QCQP, yshift=1cm] (affCon) {Affine \\ Constraints? \\ ${}$};
\node [block, below of=quadFun] (affTrans) {Remove Affine Constraint};
\node [block, below of=affCon, yshift=0.5cm, xshift=-1.7cm] (centralize) {Apply \\ Transformation $\Ivec = \M{x} - \dfrac{1}{2} \M{B}_1^{-1} \V{b}_1$};
\node [decision, below of=centralize, yshift=0.5cm] (linTerm) {Linear Terms?};
\node [decision, right of=linTerm] (constr1) {One Constraint?};
\node [block, below of=linTerm] (codeN) {Use \mtbtext{minAstBn.m}};
\node [block, right of=codeN] (code1) {Use \mtbtext{minAstB1.m}};
\node [block, left of=codeN] (codeLin1) {Use \mtbtext{minLinStB1.m}};
\node [block, left of=codeLin1] (codeLin) {Use \mtbtext{QNCQPQuadLin.m}};
\node [decision, left of=linTerm, xshift=-1.7cm] (linObj) {Linear Objective Function and One Constraint?};
\node [block, below of=codeN, yshift=1.5cm, xshift=-1.7cm] (solution) {Apply Inverse Transformations};

\path [line] (definition) -- (quadFun);
\path [line,dashed] (operators) -- (quadFun);
\path [line,dashed] (metrics) -- (operators);
\path [line,dashed] (metrics) -- (definition);
\path [line] (quadFun) -- (QCQP);
\path [line] (QCQP) -- (affCon);
\path [line] (affCon) -- node [above] {yes} (affTrans);
\path [line] (affTrans) |- (centralize);
\path [line] (affCon) -- node [right] {no} (centralize);
\path [line] (centralize) -- (linTerm);
\path [line] (linTerm) -- node [above] {yes} (linObj);
\path [line] (linObj) -- node [left] {no} (codeLin);
\path [line] (linObj) -- node [right] {yes} (codeLin1);
\path [line] (linTerm) -- node [above] {no} (constr1);
\path [line] (constr1) -- node [right] {yes} (code1);
\path [line] (constr1) -- node [left] {no} (codeN);
\path [line] (codeLin) |- (solution);
\path [line] (codeLin1) -- (solution);
\path [line] (codeN) -- (solution);
\path [line] (code1) |- (solution);

\end{tikzpicture}
\caption{Fundamental bound searching flow chart.}
\label{fig:flowChart}
\end{figure}
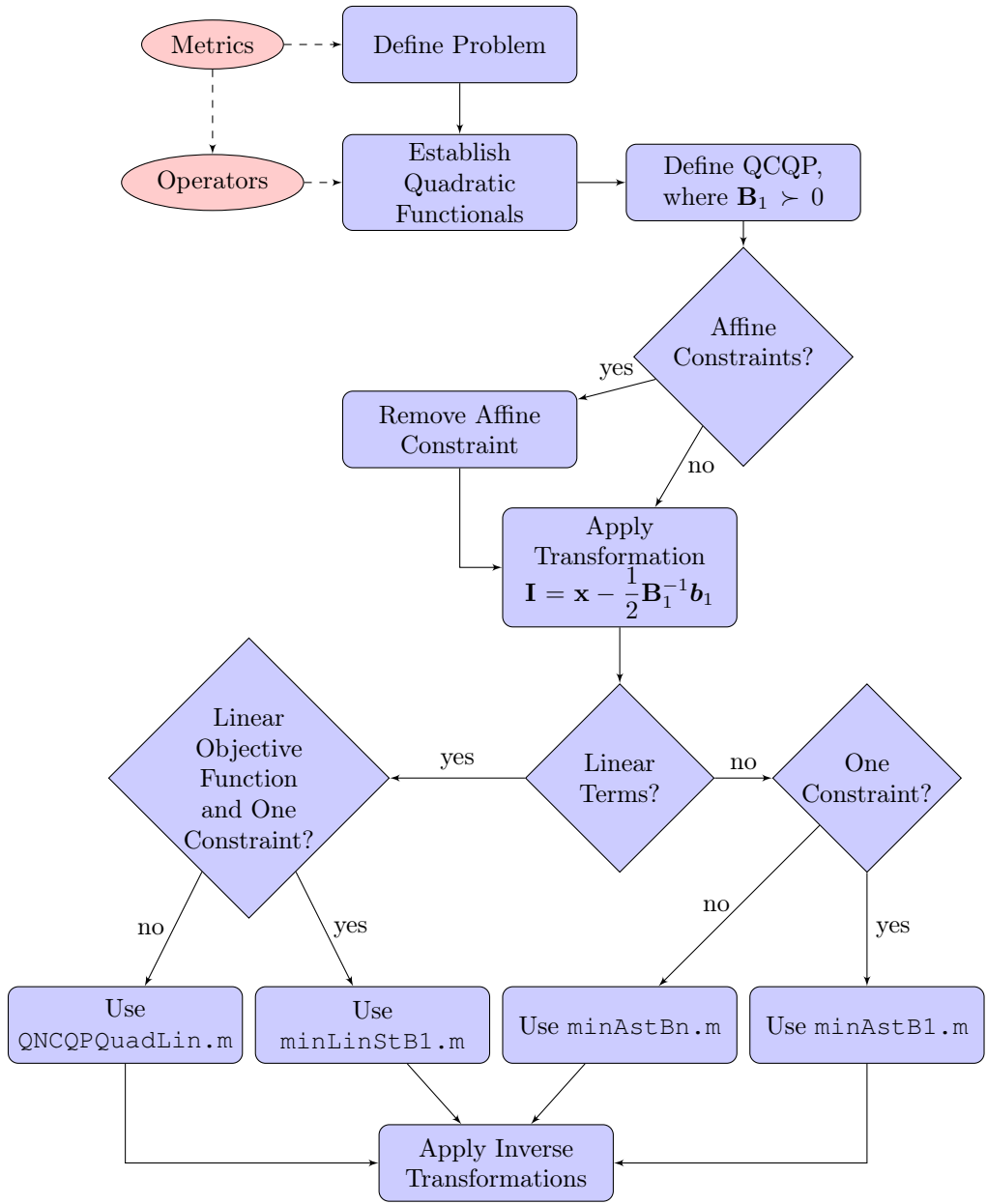

Determination of a fundamental bound using the algorithms in Appendix~\ref{ap:coreFunctions} requires specific steps to be followed and the conditions mentioned in this chapter to be fulfilled. The procedure is visualized in Fig.~\ref{fig:flowChart}.

First, it is important to decide whether the problem at hand is single- or multi-objective. The described algorithms are solely able to find a solution to \ac{QCQP} with one optimized metric and an arbitrary number of constraints, i.e., in the case of a multi-objective problem, it is solely able to find a single point of a Pareto-optimal set. Two possibilities of how a multi-objective problem can be transformed in this respect are given in Appendix~\ref{ap:pareto}.

Second, the setup of an optimization problem and its subsequent transcription into the form of a \ac{QCQP} is required, in addition to the first constraint matrix being positive definite. To simplify this step, the basic electromagnetic operators produced by \ac{MoM} are given in Appendix~\ref{ap:matricesELMAG}.

If the optimization problem contains affine constraints\footnote{The definition of controllable and uncontrollable regions (substructure bounds) is also treated by affine constraints, which is used in Sec.~\ref{sec:affine}.}, the optimization problem should be transformed so as to satisfy these constraints. The resulting optimization problem is then constrained only by quadratic functions. The transformation is explained in Appendix~\ref{ap:affineConstraints}.

The linear term appearing in the quadratic form of the first constraint should subsequently be removed by applying transformation\footnote{If this linear term is zero, the transformation~\eqref{eq:centralization} is an identity transformation.}~$\M{I} \to \M{x}$
\begin{equation}\label{eq:centralization}
    \Ivec = \M{x} - \frac{1}{2} \M{B}_1^{-1} \V{b}_1
\end{equation}
to all quadratic functionals. The transformation centers the variable space according to the first constraint, suppresses the linear term in the first constraint, and wipes out potentially hidden zero linear terms in the other constraints.

At the end of the process, all transformations that have been used, such as~\eqref{eq:centralization} or~\eqref{eq:AffTransform}, have to be restored by the application of their inverses to obtain the true optimized vector~$\M{I}$.
\section{Examples Implementation}\label{ap:exImplementation}
This Appendix supplements Section~\ref{sec:examples} with the \MATLAB implementation of the corresponding examples. The implementation is discussed briefly and focuses only on essential points. For a more detailed understanding, download and open the package with examples at the \acs{AToM} web page, section Fundamental Bounds, \url{http://antennatoolbox.com/fundamentalBounds}.

\subsection{Lower Bound on Q-factor}
The implementation of the example follows \ac{QCQP}~\eqref{eq:Q:QCQP}.

The matrices are normalized and the problem is solved by function \mtbtext{minAstBn()}, see Program~\ref{prg:minQselfRes}, which is described in Appendix~\ref{ap:coreFunctions}. The example is prepared in the file \mtbtext{exMinQselfRes} which applies the prepared function \mtbtext{minQselfRes()}.
\begin{program}
\mtbfile[51][62]{codes/minQselfRes.m}
  \caption{The core of function \mtbtext{optimProblems.minQselfRes()}. Structure \mtbtext{OP} contains the required matrices: $\Rmat$, $\Wmat$, $\Xmat = \Xmat_0 + \Xmat_\rho$. The matrices are normalized, the solution is found by function \mtbtext{solvers.QCQP.minAstBn} and the Q-factor is evaluated.}
  \label{prg:minQselfRes}
\end{program}

\subsection{Lower Bound on Dissipation Factor}
The implementation is identical to the previous case of the minimum Q-factor with only the matrices being changed, see Program~\ref{prg:minDeltaSelfRes}.
\begin{program}
\mtbfile[51][61]{codes/minDeltaSelfRes.m}
\caption{The core of function \mtbtext{optimProblems.minDeltaSelfRes()} following the same steps as Program~\ref{prg:minQselfRes}. Additionally, structure \mtbtext{OP} contains matrix $\RmatL$.}
\label{prg:minDeltaSelfRes}
\end{program}

The example is prepared in the file \mtbtext{exMinDeltaSelfRes} which applies the prepared function \mtbtext{minDeltaSelfRes()}.

\subsection{Pareto-Optimal Set of the Dissipation factor and Q-factor}
The implementation is similar to the previous examples. The major difference is a sweep (for-loop) over a convex parameter in~\eqref{eq:minQdelta} which provides a sweep over the Pareto frontier. The function \mtbtext{minAstBn()} is applied in every cycle, see Program~\ref{prg:exMinQselfResVsDelta}. 
\begin{program}
\mtbfile[50][55]{codes/exMinQselfResVsDelta.m}
\caption{Computation of one Pareto-optimal point, which is a part of the example \mtbtext{exMinQselfResVsDelta}. In each Pareto-optimal point the convex combination of positive definite matrices is normalized as a new matrix and the solver function \mtbtext{solvers.QCQP.minAstBn} is applied.}
\label{prg:exMinQselfResVsDelta}
\end{program}

\subsection{Trade-off Beween Directivity, Dissipation factor and Q-factor} \label{impl:QdeltaD}
As shown in \ac{QCQP}~\eqref{eq:minQdeltaD}, the Pareto-optimal set of points is, in this case, solved as a convex combination between the dissipation factor and the Q-factor with directivity fixed by a constraint. Due to the need of a two-dimensional sweep, all matrices are mapped to a special subspace\footnote{The chosen characteristic modes are important to the Q-factor and dissipation factor, while the singular eigenvectors enhance directivity.} in Program~\ref{prg:QdeltaDtransformMat} which decreases the complexity of computation. For the particular case used in Section~\ref{sec:examples}, the dimension of all matrices was decreased from 285 to 34 with just a small sacrifice in precision.
\begin{program}
\mtbfile[48][53]{codes/exMinQvsDeltaVsDswpSelfRes.m}
\caption{The subspace forming new basis vectors consists of characteristic modes with eigenvalues of sufficiently small magnitude, and of right singular vectors of singular value decomposition of the far-field matrix. All matrices are part of structure \mtbtext{OP}. Matrix \mtbtext{V} contains the basis vectors in its columns.}
\label{prg:QdeltaDtransformMat}
\end{program}

The subspace only consists of significant characteristic modes and the nonsingular vectors (one or two) of the radiation intensity matrix. Characteristic modes are important to the dissipation factor and the Q-factor, while the additional vector resolves the directivity.

If broadside radiation is desired, the two far-field matrices (to two broadside directions) are the same, while if the end-fire is desired, the matrices are different. One must then choose the maximum of these two function values if directivity is not included in the optimization problem, as in the case of the Pareto frontier between the dissipation factor and the Q-factor in Program~\ref{prg:QdeltaDpareto1}.
\begin{program}
\mtbfile[103][117]{codes/exMinQvsDeltaVsDswpSelfRes.m}
\caption{Computation of the dissipation and Q-factor Pareto-optimal set of points with evaluation of maximal directivity. The optimization is identical to Program~\ref{prg:exMinQselfResVsDelta}.}
\label{prg:QdeltaDpareto1}
\end{program}

The Pareto-optimal point definitions are given in Appendix~\ref{ap:pareto}, but it is only the directivity of the three optimized metrics which requires an additional choice of the optimum\footnote{For example, using a spherical shell as the current support, an additional optimization for directivity would be needed since \Quot{all directions} are equivalent.}.
    
The section of the code computing the rest of the Pareto-optimal points uses function \mtbtext{minAstBn.m}. Before the optimization procedure, the set of directivity values~$D_0$ is determined in Program~\ref{prg:QdeltaDminMaxD}.
\begin{program}
\mtbfile[126][135]{codes/exMinQvsDeltaVsDswpSelfRes.m}
\caption{The points with specified directivity are set to a multiple of $D_0$.}
\label{prg:QdeltaDminMaxD}
\end{program}

For a given directivity, the Pareto-optimal set of points of the dissipation factor and the Q-factor is found in Program~\ref{prg:QdeltaDpareto}.
\begin{program}
\mtbfile[138][154]{codes/exMinQvsDeltaVsDswpSelfRes.m}
\mtbfile[177][183]{codes/exMinQvsDeltaVsDswpSelfRes.m}
\caption{Computation of the Pareto-optimal points in all three metrics simultaneously. The procedure is similar to the Q-factor and dissipation factor Pareto-optimal points in Program~\ref{prg:exMinQselfResVsDelta}. The only difference is that the constraint for directivity is added and represented by matrix \mtbtext{matD}.}
\label{prg:QdeltaDpareto}
\end{program}
The algorithm starts with the lowest directivity found on the $Q-\delta$ Pareto frontier given by the example in Section~\ref{sec:minQdelta}. If the set directivity is lower than the directivity computed on the $Q-\delta$ Pareto frontier, the point is skipped.

\subsection{Substructure Bounds}
\label{impl:SubStructure}
The implementation of the example with substructure bounds shown in Section~\ref{sec:affine} is quite different from the previous ones. It contains affine constraints treated by Appendix~\ref{ap:affineConstraints} and applies another optimization solver.

For the purpose of this example, the mesh grid is generated by the function \mtbtext{models.utilities.meshPublic.pixelGridToOrthoMesh()} from \ac{AToM} and positioned after being resized to the chosen dimensions in Program~\ref{prg:affMeshGen}.
\begin{program}
\mtbfile[24][56]{codes/exMaxPaCUZpowerConstMoM2D.m}
\caption{Mesh generation from the pixel grid applies function \mtbtext{models.utilities.meshPublic.pixelGridToOrthoMesh()} from \ac{AToM}. The edges of the patch and ground are rescaled to the given dimensions. The meshes are joined and the structure \mtbtext{mesh} is generated.}
\label{prg:affMeshGen}
\end{program}
The incident planewave excites the whole structure from a given direction in Program~\ref{prg:affExcitation}.
\begin{program}
\mtbfile[70][75]{codes/exMaxPaCUZpowerConstMoM2D.m}
\caption{Setting excitation by a linearly polarized plane wave.}
\label{prg:affExcitation}
\end{program}
The setting of the controllable and uncontrollable parts of the structure is done by the function \mtbtext{controllableRegion.defineControllableRegion()} from the package and the indices are extracted in Program~\ref{prg:affControllable}.
\begin{program}
\mtbfile[77][89]{codes/exMaxPaCUZpowerConstMoM2D.m}
\caption{Setting of the controllable and uncontrollable regions.}
\label{prg:affControllable}
\end{program}
Then, in a cycle with a frequency sweep, the required operators are computed in Program~\ref{prg:affOperators},
\begin{program}
\mtbfile[138][149]{codes/exMaxPaCUZpowerConstMoM2D.m}
\caption{Operators in structure \mtbtext{OP} are computed from the data in the structure \mtbtext{mesh}.}
\label{prg:affOperators}
\end{program}
the affine constraints are set in Program~\ref{prg:affConstraints},
\begin{program}
\mtbfile[151][172]{codes/exMaxPaCUZpowerConstMoM2D.m}
\caption{Affine constraints are stored in two variables (\mtbtext{OP.Aaff} , \mtbtext{OP.aAff}) as rows. They are set to satisfy $\Zmat_\T{uc} \Ivec - \Vvec_\T{u} = \bm{0}$ in the uncontrollable region and to ensure zeros in the scattering pattern.}
\label{prg:affConstraints}
\end{program}
and the optimization problem is implemented in the function \mtbtext{optimProblems.minMaxPaPsPeZpowerConst()}, see Program~\ref{prg:affSolution}.
\begin{program}
\mtbfile[176][182]{codes/exMaxPaCUZpowerConstMoM2D.m}
\caption{The optimization problem is solved in the function \mtbtext{controllableRegion.transformQuadFormAff()}, which uses the solver function designed for problems with linear terms.}
\label{prg:affSolution}
\end{program}
The function employs \mtbtext{controllableRegion.transformQuadFormAff()} to treat the affine constraint and \mtbtext{solvers.QCQP.QNCQPQuadLin()} to solve the optimization problem.
\section{Multi-Objective Optimization and Pareto-Optimal Sets} \label{ap:pareto}

Multi-objective optimization can be formulated as~\cite{1978CohonMultiobjectiveProgrammingAndPlanning,Deb_MultiOOusingEA,2008EichfelderAdaptiveScalarizationMethodsInMultiobjectiveOptimization}

\begin{equation}
    \min \limits_{\Ivec} \quad \left\{ f_1(\Ivec), f_2(\Ivec), \ldots, f_p(\Ivec) \right\}
\end{equation}
where $p \geq 2$ is the number of objectives and arbitrary constraints could also be included. A solution to the multi-objective optimization is commonly assumed in the form of a Pareto-optimal set, which is a set of~$\Ivec$ such that a decrease in any objective function cannot be made without an increase in another function. One possibility of evaluating the Pareto-optimal set, which is used in this text, is to define a function
\begin{equation}
    f(\bm{c}, \bm{w}, \Ivec) = \sum \limits_{i=1}^p c_i w_i f_i(\Ivec), \ \T{where} \ \sum \limits_{i=1}^p c_i = 1, \ c_i \geq 0, \ w_i > 0
    \label{eq:MultiObjConvexComb}
\end{equation}
which is a convex combination of objective functions weighted by arbitrary positive weights~$w_i$. The set of points
\begin{equation}
    \hat{\Ivec}(\bm{c}) = \argmin \limits_{\Ivec} f(\bm{c}, \bm{w}, \Ivec),
\end{equation}
is Pareto-optimal for arbitrary fixed weights~$\bm{w}$ and all $c_i > 0$.

If $c_k = 1$, the other coefficients $c$ are zero and
\begin{equation}
\hat{f}_k = \min \limits_{\Ivec} \quad f_k(\Ivec)
\end{equation}
then 
\begin{align}
    \min \limits_{\Ivec} \quad & \sum \limits_{i \neq k} f_i(\Ivec) \\
\T{s.t.} \quad & f_k (\Ivec) = \hat{f}_k
\end{align}
is Pareto-optimal for all $0<c_i<1, \ \forall i \neq k$. If any $c_k = 0$,
\begin{equation}
    \hat{\Ivec} = \argmin \limits_{\Ivec} \ \sum \limits_{i \neq k} f_i(\Ivec),
\end{equation}
then
\begin{align}
    \min \limits_{\Ivec} \ & f_k(\Ivec) \\
    \T{s.t.} \quad & f_i (\Ivec) = f_i (\hat{\Ivec}), \ \forall i \neq k
\end{align}
is Pareto-optimal with the previous assumptions and recurrence.

The linear combination~\eqref{eq:MultiObjConvexComb} is not able to cover~\cite{1978CohonMultiobjectiveProgrammingAndPlanning} linear trade-offs between metrics in the Pareto-optimal sets. Such cases can be treated by optimizing one of the metrics only while fixing the others by constraints, i.e.,
\begin{align}
    \min \limits_{\Ivec} \quad & f_k(\Ivec) \\
    \T{s.t.} \quad & f_i (\Ivec) = \Tilde{f}_i, \ \forall i \neq k,
\end{align}
where $\Tilde{f}_i$ is a Pareto-optimal point according to metrics $f_i, \ \forall i \neq k$.

\section{Removal of an Affine Constraint} \label{ap:affineConstraints}
Assume an affine constraint
\begin{equation}
\label{eq:affConst}
    \M{A} \M{I} + \M{a} = 0,
\end{equation}
where~$\M{A}$ is a complex rectangular matrix of size~$M \times N$ with~$M < N$. This appendix shows a basis transformation
\begin{equation}
\label{eq:AffTransform}
    \M{I} = \M{\M{t}} + \M{\M{T}} \M{x}
\end{equation}
that removes this constraint from the optimization problem. 

In order to obtain vector~$\M{\M{t}}$ and matrix~$\M{\M{T}}$, matrix~$\M{A}$ is decomposed via singular value decomposition~\cite{GolubVanLoan_MatrixComputations} as
\begin{equation}
    \M{A} = \mqty[\M{U}_1^\T{L} & \M{U}_1^\T{R}] \mqty[\M{\sigma} & \M{0} \\ \M{0} & \M{0}] \mqty[\M{U}_2^\T{L} & \M{U}_2^\T{R}]^\herm,
\end{equation}
where matrices~$\M{U}$ are unitary and~$\M{\sigma}$ contains non-zero singular values or singular values that were above the user-defined threshold. Substituting into~\eqref{eq:affConst} it is possible to see that transformation~\eqref{eq:AffTransform} with
\begin{equation}
\begin{aligned}
    \M{\M{t}} &= - \M{U}_2^\T{L} \T{diag} \left( \dfrac{1}{\sigma_n} \right) \left( \M{U}_1^\T{L} \right)^\herm \\
    \M{\M{T}} &= \M{U}_2^\T{R}
\end{aligned}    
\end{equation}
is the solution (or least-squares approximation) to~\eqref{eq:affConst} for arbitrary vector~$\M{x}$.

Knowing transformation~\eqref{eq:AffTransform}, the affine constraint~\eqref{eq:affConst} can be removed from the optimization problem by transforming all its quadratic forms~\eqref{eq:guadFcnl} as
\begin{equation}
    f \left( \M{x} \right) = \M{x}^\herm \widetilde{\M{A}} \M{x} + \RE[\M{x}^\herm \widetilde{\M{a}}] + \widetilde{\alpha},
\end{equation}
where
\begin{align}
    \widetilde{\M{A}} &= \M{T}^\herm \M{A} \M{T}, \\
    \widetilde{\M{a}} &= \M{T}^\herm \left(2 \M{A} \M{t} + \M{a} \right), \\
    \widetilde{\alpha} &= \RE \left[ \M{t}^\herm \left(\M{A} \M{t} + \M{a} \right) + \alpha \right].
\end{align}
This functionality is provided by the \MATLAB functions:
\begin{itemize}
\item \mtbtext{controllableRegion.createTransformAffine()}, \item \mtbtext{controllableRegion.transformQuadFormAff()}
\end{itemize}
from the package.

\section{Implementation of QCQP Solvers} \label{ap:coreFunctions}
All examples mentioned in section~\ref{sec:examples} use one of the in-house solvers contained in the name space \mtbtext{+solvers\+QCQP} which are prepared to solve (or to set a lower bound to) arbitrary \ac{QCQP} using a dual formulation. The solvers follow the nomenclature of Appendix~\ref{ap:QCQP}. If the values of the primal and the dual problem are not the same, a dual gap occurs~\cite{BoydVandenberghe_ConvexOptimization} and a warning report is given.

There are four different solvers, each dedicated to the appropriate form of \ac{QCQP}. Functions \mtbtext{minLinStB1} and \mtbtext{minAstB1} are applicable to quadratic programs with one purely quadratic constraint and a purely linear or purely quadratic objective function. These functions are specialized, simple to read and are not detailed in this appendix. The other two solvers \mtbtext{QNCQPQuadLin} and \mtbtext{minAstBn} are generic and are detailed in this appendix.

\subsection{Quadratic Objective Function and Quadratic Constraints with Linear Terms (\mtbtext{QNCQPQuadLin})}
This function solves full \ac{QCQP} of the form \eqref{eq:genProb}--\eqref{eq:genProbConstr} with at least one linear term, i.e., with dual function~\eqref{eq:dualLin}. The solver begins with the Cholesky factorization of the positive definite matrix~$\M{B}_1$ in Program~\ref{prg:cholesky} to decrease the complexity of the subsequent eigenvalue decompositions.
\begin{program}
\mtbfile[49][49]{codes/QNCQPQuadLin.m}
\caption{Cholesky decomposition of the positive definite matrix.}
\label{prg:cholesky}
\end{program}
Initial Lagrange multipliers are compared to the maximal one in Program~\ref{prg:lambdaLeadMax} which ensures the positive definiteness of the Hessian matrix.
\begin{program}
\mtbfile[338][339]{codes/QNCQPQuadLin.m}
\caption{Maximal Lagrange multiplier to stay in the positive definite region.}
\label{prg:lambdaLeadMax}
\end{program}
The initial solution and value of the dual function are computed in Program~\ref{prg:initialDualFun}.
\begin{program}
\mtbfile[82][86]{codes/QNCQPQuadLin.m}
\caption{Computation of the initial value of the dual function.}
\label{prg:initialDualFun}
\end{program}
Newton's method is used to maximize the dual function. Newton's step in Lagrange multipliers is given by
\begin{equation}
\M{p} = - \M{H}_\T{d}^{-1} \V{g}_\T{d},
\end{equation}
where
\begin{equation}
    H_\T{d}^{ij} = 2 \RE{\left[\left(\M{B}_i \V{x} + \frac{\M{b}_i}{2}\right)^\herm \M{H}^{-1} \left(\M{B}_j \M{x} + \frac{\M{b}_j}{2}\right)\right]}
\end{equation}
is an element of the dual function Hessian matrix for the actual vector~$\M{x}$ and
\begin{equation}
    g_\T{d}^{i} = - \RE{\left[\left(\M{B}_i \M{x} + \M{b}_i\right)^\herm \M{x} \right]} - \beta_i
\end{equation}
is an element of the dual function gradient at the same point in Program~\ref{prg:proposedShift}.
\begin{program}
\mtbfile[142][157]{codes/QNCQPQuadLin.m}
\caption{Computation of Newton's shift based on the gradient and Hessian matrix of the dual function.}
\label{prg:proposedShift}
\end{program}
The length of the step is controlled and reduced in Program~\ref{prg:controlledShift} by a coefficient which ensures the positive definiteness of the new Hessian matrix.
\begin{program}
\mtbfile[207][207]{codes/QNCQPQuadLin.m}
\caption{Newton's shift modified in length by a positive constant \mtbtext{alpha} which enforces the new set of Lagrange multipliers to stay in the positive definite region of the dual function.}
\label{prg:controlledShift}
\end{program}
The dual function is evaluated in the next Newton's method step as in Program~\ref{prg:initialDualFun}.
The above procedure is repeated until the relative error between the actual value and the last value of the dual function is smaller than a predefined toleration or the maximum of the allowed iterations is reached.

\subsection{Quadratic Objective Function and two or more Quadratic Constraints, all without Linear Terms (\mtbtext{minAstBn})}
A purely quadratic program with more than one purely quadratic constraint is solved by the numerical maximization of the dual function~\eqref{eq:dual}.

\begin{program}
\mtbfile[58][59]{codes/minAstBn.m}
\caption{Matrix transformation and symmetrization.}
\label{prg:matTransformSym}
\end{program}
To increase the efficiency of the code, all matrices are transformed
\begin{equation}
    \Tilde{\M{M}} = \left( \M{R}^\herm \right)^{-1} \M{M} \M{R}^{-1},
\end{equation}
where $\M{R}$ is the Cholesky factor to matrix $\M{B}_1$,
\begin{equation}
    \M{B}_1 = \M{R}^\herm \M{R}
\end{equation}
in Program~\ref{prg:matTransformSym}. Then, the \ac{GEP}~\eqref{eq:gep} is reduced to the ordinary eigenvalue problem
\begin{equation} \label{eq:ep}
    \left(\Tilde{\M{A}}-\sum \limits_{i=2}^m \mu_i \Tilde{\M{B}}_i \right)
    \M{Y} =
    \lambda \M{Y}, \qquad \Tilde{\Ivec} = \M{R}^{-1} \M{Y}.
\end{equation}
The enhancement is used in the dual problem implementation in Program~\ref{prg:dualProblem}.
\begin{program}
\mtbfile[163][184]{codes/minAstBn.m}
\caption{Dual problem formulation. The matrices are summed together with the given Lagrange multipliers. The desired eigenvalue is found (with numerical problem treatment), the dual function is evaluated and saved in output variable \mtbtext{dP}, the minimal eigenvalue is shared from this nested function by variable \mtbtext{dMin}.}
\label{prg:dualProblem}
\end{program}
Dual function is minimized numerically by the \MATLAB function \mtbtext{fminsearch()} in Program~\ref{prg:optimDualP}.
\begin{program}
\mtbfile[90][90]{codes/minAstBn.m}
\caption{Dual problem maximization.}
\label{prg:optimDualP}
\end{program}
If the eigenvalue is multiple, the symmetry treatment is required to match the value of the primary problem to the value of the dual problem, i.e., to close the fictitious dual gap~\cite{2021_Capek_TAP_SymmetriesAndBounds}. The associated eigenvectors are linearly combined to minimize the norm of a vector whose elements are the values of the constraints in Program~\ref{prg:symmetryTreatment}.
\begin{program}
\mtbfile[186][201]{codes/minAstBn.m}
\mtbfile[220][228]{codes/minAstBn.m}
\mtbfile[230][244]{codes/minAstBn.m}
\caption{Symmetry treatment and constraint fitting is done by minimizing the norm of error in the constraints.}
\label{prg:symmetryTreatment}
\end{program}

\bibliographystyle{ieeetr}
\bibliography{references.bib,addRef.bib}
\end{document}